\documentclass[article,nofootinbib,superscriptaddress]{revtex4}

\usepackage{amsfonts}
\usepackage{bm}
\usepackage{color}
\usepackage{graphicx}
\usepackage{amsmath}
\usepackage{amsfonts}
\usepackage{amssymb}
\usepackage{color}
\usepackage{epsf}
\usepackage[a4paper]{geometry} 
\usepackage[bookmarksdepth=2]{hyperref}

\newcommand{\be}{\begin{equation}}
\newcommand{\ee}{\end{equation}}
\newcommand{\ba}{\begin{eqnarray}}
\newcommand{\ea}{\end{eqnarray}}

\newcommand{\wj}[6]{\left(
                           \begin{array}{ccc}
        \! #1\! & #2\!  & #3\!  \\
        \! #4\! & #5\!  & #6\!
                           \end{array}
                   \right)}

\newcommand{\svm}{(\langle\sigma v\rangle/m_\chi)}

\newcommand{\jl}[3]{j_{#1}[k_{#2}(\eta_0-\eta_{#3})]}

\newcommand{\no}{\nonumber}

\begin{document}

\vspace*{-30mm}

\title{\boldmath Perturbed Recombination from Dark Matter Annihilation}

\author{Cora Dvorkin}\email{cdvorkin@ias.edu}
\author{Kfir Blum}\email{kblum@ias.edu}
\author{Matias Zaldarriaga}\email{matiasz@ias.edu}
\affiliation{School of Natural Sciences, Institute for Advanced Study, Princeton, New Jersey 08540, USA}

\vspace*{1cm}

\begin{abstract} 
We show that dark matter annihilation around the time of recombination can lead to growing ionization fraction perturbations, that track the linear collapse of matter over-densities. This amplifies small scale cosmological perturbations to the free electron density by a significant amount compared to the usual acoustic oscillations. Electron density perturbations distort the CMB, inducing secondary non-gaussianity. We calculate the CMB bispectrum from recombination, that is marginally observable by Planck. Even though electron perturbations can be markedly boosted compared with the Standard Model prediction, the dark matter effect in the CMB bispectrum turns out to be small and will be difficult to disentangle from the standard model in the foreseeable future. 
\end{abstract}

\maketitle

\section{Introduction}
\label{sec:intro}

Galaxy and galaxy cluster dynamics and cosmological observations seem to imply that five out of six parts in mass of all matter in the Universe is composed of dark matter (DM), that is not accounted for by the Standard Model of particles. The particle nature of DM is one of the most intriguing puzzles of our time. Many efforts are invested in trying to solve this puzzle at direct and indirect detection experiments. It is important to identify astrophysical and cosmological processes where the particle interactions of DM, rather than its gravitational pull alone, may be of relevance. 
In this paper, we discuss a cosmological observable where DM interactions can modify appreciably the Standard Model prediction: linear perturbations to the ionization fraction of hydrogen. 

To summarize our main findings: DM annihilation can significantly change the evolution of linear cosmological perturbations to the free electron density, at and after the last scattering epoch of the Cosmic Microwave Background (CMB). Consistent with all current constraints, the effect can be as large as an $\mathcal{O}(10)$ enhancement for perturbations on small scales. Of course, it is not enough for DM effects to be large. For us to learn about it, the effect must also be visible. In the current paper, we looked for observable imprints of the electron perturbations in CMB non-gaussiainity. To no avail: even though electron perturbations can be markedly boosted, the main boost occurs slightly after last scattering, and on scales below the Silk damping scale, and so the non-gaussianity signal is small. In the rest of this introduction we expand on our motivations and lay out the structure of the paper.

As is well known, dark matter annihilation or decay could modify the ionization history of the universe, giving rise to excess Thomson scattering compared to the Standard Model prediction~\cite{Chen:2003gz,Zhang:2006fr,Belikov:2009qx,Galli:2009zc,Slatyer:2009yq,Cirelli:2009bb,Natarajan:2010dc,Galli:2011rz,Finkbeiner:2011dx,Hutsi:2011vx,Natarajan:2012ry,Giesen:2012rp,Evoli:2012qh}. This extra scattering damps power in small scale CMB temperature anisotropies and adds power in polarization. For DM annihilation, CMB constraints apply to the parameter combination $\svm$, where $\langle\sigma v\rangle$ is the velocity-weighted annihilation cross section and $m_\chi$ is the DM mass. For simple thermal freezeout models, the cross section is fixed $\langle\sigma v\rangle\sim3\times10^{-26}$~cm$^3$s$^{-1}$GeV$^{-1}$ and current CMB constraints based on precision measurements ~\cite{Hinshaw:2012fq,Dunkley:2013vu,Hou:2012xq} become important for DM masses below and of order 10~GeV. Planck data expected in the near future will either provide a detection or tighten the constraints. 

All of the CMB constraints in the current literature apply to temperature and polarization two-point correlation functions, or power spectra. Existing analyses usually consider the homogeneous ionization history, where the free electron density is taken to be a function of time only, $n_e=n_e(t)$. 
In some analyses (e.g., recently, \cite{Giesen:2012rp,Slatyer:2012yq}), late time ($z\lesssim30$) annihilation in non-linear halos is included but only in terms of a (model-dependent) boost to the smooth component. 

In contrast, our interest in the current paper is with linear cosmological perturbations to the electron density, characterized by $\delta_e$ such that $n_e\to n_e(t)\left(1+\delta_e(\vec x,t)\right)$. Power spectra are insensitive to these fluctuations, because in two point functions they enter at fourth (4th) order in the primordial curvature perturbations $\xi\sim10^{-5}$. Thus they cannot compete with the leading quadratic contributions. The leading observable where $\delta_e$ may play a role is CMB non-gaussianity, in particular the three-point function or bispectrum. Many inflationary models predict a very small primordial bispectrum so the first non-zero contribution may be due to deviations from linear evolution, that can be described via second order cosmological perturbation theory. A finite first order $\delta_e$ produces second-order CMB inhomogeneities, transforming into finite anisotropy bispectrum.

The paper is organized as follows. In \S\ref{sec:inhom_recomb} we calculate electron density perturbations. First, in \S\ref{ssec:homde} we review the standard recombination model and show that DM can make a sizable impact on the homogeneous hydrogen ionization fraction $x_e$ after last scattering, that may in fact be dominated by DM annihilation. As is well known, an $\mathcal{O}(1)$ relative correction here is allowed experimentally, because (i) the residual ionization $x_e$ after last scattering is small anyway, and (ii) in the temperature power spectrum, the best measured CMB observable, excess Thomson optical depth is partially degenerate with the normalization $A_s$ and tilt $n_s$ of the primordial curvature fluctuations. 

In \S\ref{ssec:inhomde} we move on to cosmological perturbations. We build on the analysis of~\cite{Senatore:2008vi} and give a semi-analytical derivation of the perturbation $\delta_e$ that applies at high redshift $z\gtrsim700$, relevant for CMB studies. It was realized in \cite{Novosyadlyj:2006fw,Venhlovska:2008uc} that at early times electron density perturbations follow an amplified copy of the baryon acoustic oscillations, with amplification factor $\sim5$ corresponding to ionization waves. Extending the analysis to include DM annihilation we find a growing, non-oscillating, ionization mode that tracks the DM perturbations. The main result of this paper is that on small scales, this growing mode can boost $\delta_e$ by more than an order of magnitude compared to the Standard Model prediction, with peak amplification right after last scattering. Our calculations generalize earlier work that focused on later times long after recombination, see e.g.~\cite{Furlanetto:2006wp}. 

In \S\ref{sec: sec_order_anisotropies} we consider non-gaussianity. Several analytical and numerical studies have shown that the bispectrum from recombination is relevant for Planck and should be accounted for when searching for primordial non-gaussianity~\cite{Senatore:2008wk,Khatri:2008kb,Khatri:2009ja,Pitrou:2010sn,Huang:2012ub,Su:2012gt,Pettinari:2013he}. The leading sources appear to be second order metric  and first order electron perturbations, inducing second order radiation terms. 
Refs.~\cite{Senatore:2008wk,Khatri:2008kb,Khatri:2009ja} found the bispectrum induced by $\delta_e$ may be marginally observable by Planck. An order of magnitude amplification by DM annihilation then looks naively quite promising; we therefore compute the bispectrum induced by $\delta_e$. In doing so, we have found the current literature lacking, specifically when it comes to perturbations on small scales. Our treatment of this problem will be reported separately in~\cite{BDZ}. 

Our analysis shows that unfortunately, DM annihilation has little impact on the recombination bispectrum. The main reasons for this are: (i) the  amplification to electron perturbations peaks immediately after CMB last scattering -- largely missing the visibility window and hitting the early Dark Ages, instead; (ii) the DM effect rises on small scales below the Silk damping scale; (iii) in general, short wave electron fluctuations cannot affect long wave photon modes. This reduces the effect on squeezed triangles, where much of the signal-to-noise for the bispectrum is.

DM annihilation can affect the evolution of matter temperature perturbations during the  cosmic dark ages. While we do not pursue this avenue here, a natural means to try and detect the effect in the future would be through observations of 21~cm absorption~\cite{Furlanetto:2006wp,Valdes:2007cu,Cumberbatch:2008rh,Finkbeiner:2008gw,Natarajan:2009bm,Valdes:2012zv}. Our calculation of the temperature perturbations extends previous analyses by properly accounting for the early initial conditions from the time of recombination.  

We conclude in \S\ref{sec:conc}. 
In App.~\ref{app:deposition} we discuss the implications of non-local energy deposition by DM annihilation. 

Throughout this paper we work with the following fiducial WMAP $7$-year \cite{Komatsu:2010fb} cosmology: $\Omega_bh^2=0.0226$, $\Omega_{DM}h^2=0.112$, $h=0.704$, $\Omega_K=0$, $\tau=0.087$, $A_s=2.16\times10^{-9}$, $n_s=0.963$, with $k_p=0.05$ Mpc$^{-1}$.

\section{Dark matter annihilation effects in the recombination history}
\label{sec:inhom_recomb}

In this section we compute the cosmological electron density to first order in perturbation theory. We show that DM annihilation can cause a growing ionization mode, beginning around the time of recombination. This growing mode can boost  electron perturbations by an order of magnitude compared to the case without DM annihilation. We start by a brief review of the homogeneous calculation and then move on to the perturbation analysis.

\subsection{Homogeneous calculation}\label{ssec:homde}

We follow the standard Peebles three-level atom formalism~\cite{1968ApJPeebles,Zeldovich,Seager:1999bc,Seager:1999km} and neglect helium ionization. The homogeneous free electron density is found by solving an effective Boltzmann equation,
\be\label{eq:3lev}\frac{\partial n_e}{\partial t}+3Hn_e=Q_e,\ee
with the DM ionization rate $I_\chi$ included in the collision term
\be Q_e=\left(\beta_He^{-\epsilon_{12}/T_R}(n_H-n_e)-\alpha_Hn_e^2\right)C_H+I_\chi.\ee
Here, the type-B recombination and ionization coefficients, $\alpha_H(T_M)$ and $\beta_H(T_R)$, are given in~\cite{Seager:1999bc,Seager:1999km}; $\epsilon_{12}=10.2$~eV denotes the first excitation energy of hydrogen; $T_M$ is the kinetic matter temperature; and $n_H$ is the total number density of hydrogen, ionized and neutral. 
The factor $C_H$ denotes the probability of an $n=2$ hydrogen atom to relax to the ground state  before being photoionized and without exciting an adjacent ground state atom. It is given by
\be\label{eq:CH} C_H={1+K_H\Lambda_H (n_H-n_e)\over 1+K_H(\Lambda_H+\beta_H)(n_H-n_e)},\ee
where $\Lambda_H\approx8.3$~Hz is the two-photon $2s\to1s$ transition rate and the $L_\alpha$ redshifting rate, $K_H^{-1}(z)=(8\pi H(z)/\lambda_{\alpha}^3)$, is described in~\cite{Seager:1999bc,Seager:1999km}.

The DM ionization term is given by
\be\label{eq:dminj} I_\chi=
\frac{\dot u}{a^4\,\epsilon_H}\,C_{ion}.\ee
The quantity $\left(\dot u/a^4\,\epsilon_H\right)$ denotes the proper rate per unit volume at which energy from DM annihilation is absorbed in the plasma, measured in units of the hydrogen ionization energy $\epsilon_H=13.6$~eV; we will return to this quantity shortly. The factor $C_{ion}$ encodes the fraction of the absorbed energy which goes to ionizing the plasma. We use a crude parametrization of the partitioning of the absorbed energy between direct ionization, atomic excitation, and heating~\cite{Chen:2003gz}: 
\be\label{eq:Cion} C_{ion}=\left(\frac{n_H-n_e}{3n_H}\right)\left(1+\frac{4}{3}\left(1-C_H\right)\right).\ee
We stress that Eq.~(\ref{eq:Cion}) is a rough estimate. A more careful account of the energy partitioning deserves further study~\cite{Valdes:2009cq,Evoli:2012zz}, but is beyond the scope of this paper.

We now discuss $\left(\dot u/a^4\right)$, the proper rate per unit volume at which energy from DM annihilation is absorbed in the plasma. For later convenience, where possible we will switch to work with conformal time $d\eta=(dt/a)$. Thus here and in what follows an over-dot represents derivative with respect to $\eta$. We will also use comoving coordinates $a\,x=x_{\rm proper}$. Note however that unless stated otherwise, we still use proper particle densities, e.g. $n_e$ denotes the free electron density per unit proper volume, etc.

Jumping ahead of ourselves for a minute by including spatial inhomogeneity, the comoving power density injected into the plasma is given by\footnote{For concreteness, we assumed Majorana DM and neglected possible time dependence in $\langle\sigma v\rangle$. We remind that $\rho_\chi$ here still refers to the proper -- not comoving -- DM density. In the homogeneous limit it is given by $a^3\rho_\chi(\eta)=a'^3\rho_\chi(\eta')$.}
\be\dot u_{inj}(\vec x,\eta)=a^4(\eta)\,\frac{\langle\sigma v\rangle}{m_\chi}\rho_\chi^2(\vec x,\eta).\ee
The energy absorption rate is, in general, different from the injected power. At the epoch of interest, namely during and after recombination, particles coming out of an annihilation event can propagate over non-negligible time before their energy is absorbed by the plasma. The propagation time and distance depend on the particle type, initial energy, and time of injection, with final states of relevance being photons, electrons, and protons (with neutrinos trivially escaping indefinitely). The local rate of energy absorption is then given by folding the injection rate with some distribution, $\mathcal{F}$, specifying the propagation of the annihilation products,
\be\label{eq:smear1} \dot u(\vec x,\eta)=\int_0^\eta d\eta'\int d^3x'\mathcal{F}\left(\vec x,\vec x',\eta,\eta'\right)\dot u_{inj}(\vec x',\eta').\ee

We will have more to say about the quantity $\mathcal{F}$ in App.~\ref{app:deposition}. For now, returning to the homogeneous calculation, the spatial integral in Eq.~(\ref{eq:smear1}) goes away and one is left with
\be\label{eq:smearhom}\dot{u}(\eta)=\int_0^\eta d\eta'\,\dot{u}_{inj}(\eta')\,f_{dep}\left(\eta,\eta'\right),\ee
where
\be f_{dep}\left(\eta,\eta'\right)=\frac{1}{\epsilon_{inj}}\frac{\partial\epsilon}{\partial\eta}=\int d^3x'\mathcal{F}\left(\vec x,\vec x',\eta,\eta'\right)\ee
describes the amount of energy absorbed by the plasma at time $\eta$ per interval $d\eta$, after injecting initial energy $\epsilon_{inj}$ at the annihilation time $\eta'$. The time integral in Eq.~(\ref{eq:smearhom}) can be factored out as~\cite{Padmanabhan:2005es,Slatyer:2009yq,Slatyer:2012yq}, 
\be\dot{u}(\eta)=\dot{u}_{inj}(\eta)f(\eta).\ee
For time-independent velocity-weighted annihilation cross section, 
\be\label{eq:f} f(\eta)=\int_0^\eta d\eta'\,\frac{a^2(\eta)}{a^2(\eta')}\,f_{dep}\left(\eta,\eta'\right).\ee

In App.~\ref{app:deposition} we discuss $f(\eta)$ in the context of concrete particle physics model examples. These examples illustrate the sensitivity of $f(\eta)$ to model details. Nevertheless, it is a reasonable approximation to take $f(\eta)$ as constant, $f$, over the  time scale of recombination; for garden variety standard model final states, $f$ ranges  between 0.3 to 1~\cite{Slatyer:2009yq,Slatyer:2012yq}. Thus in this paper, as a rule, we simply absorb $f$ into the definition of $\langle\sigma v\rangle$. 

\begin{figure}[!t]\begin{center}
\includegraphics[width=0.475\textwidth]{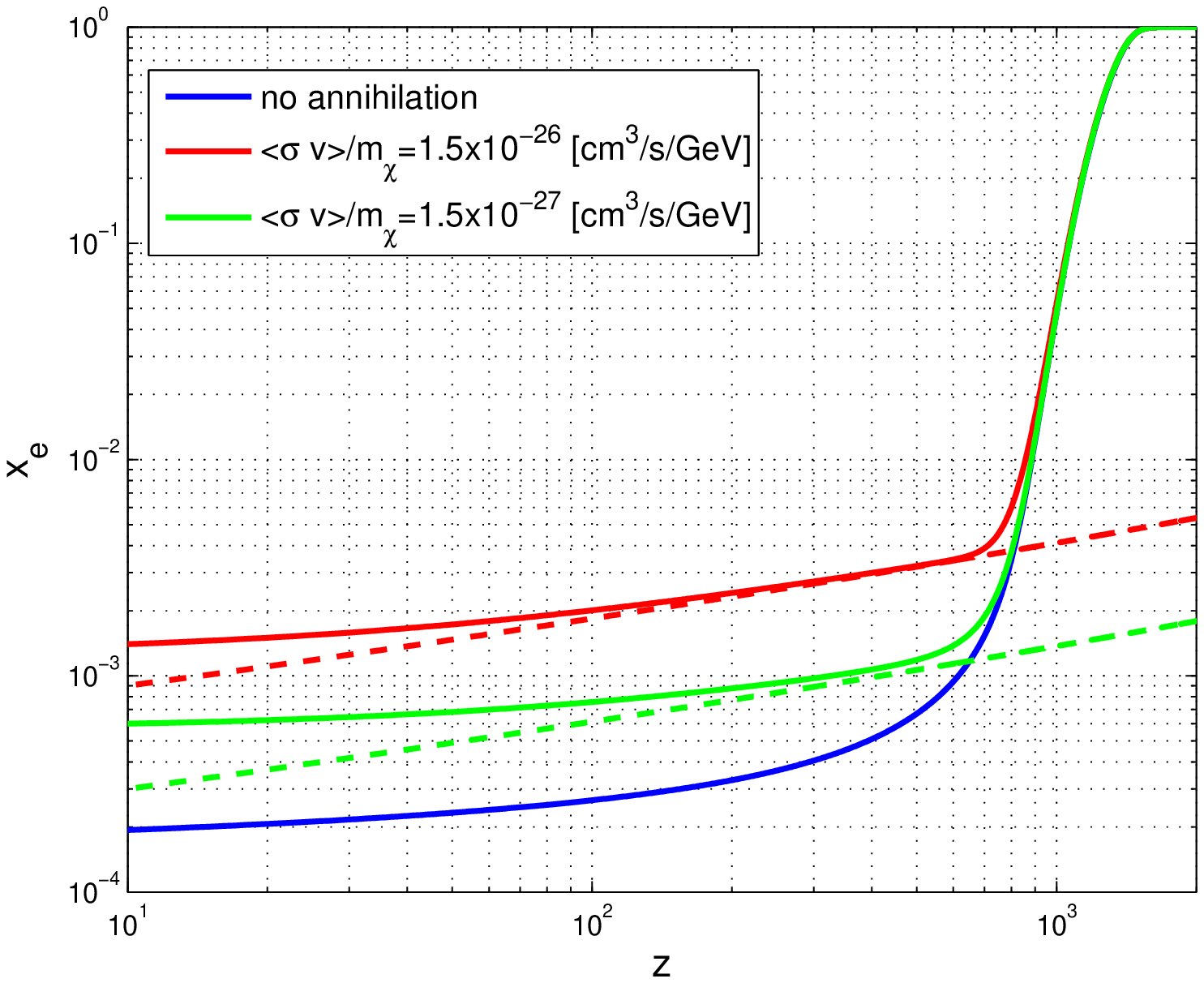}\quad
\includegraphics[width=0.475\textwidth]{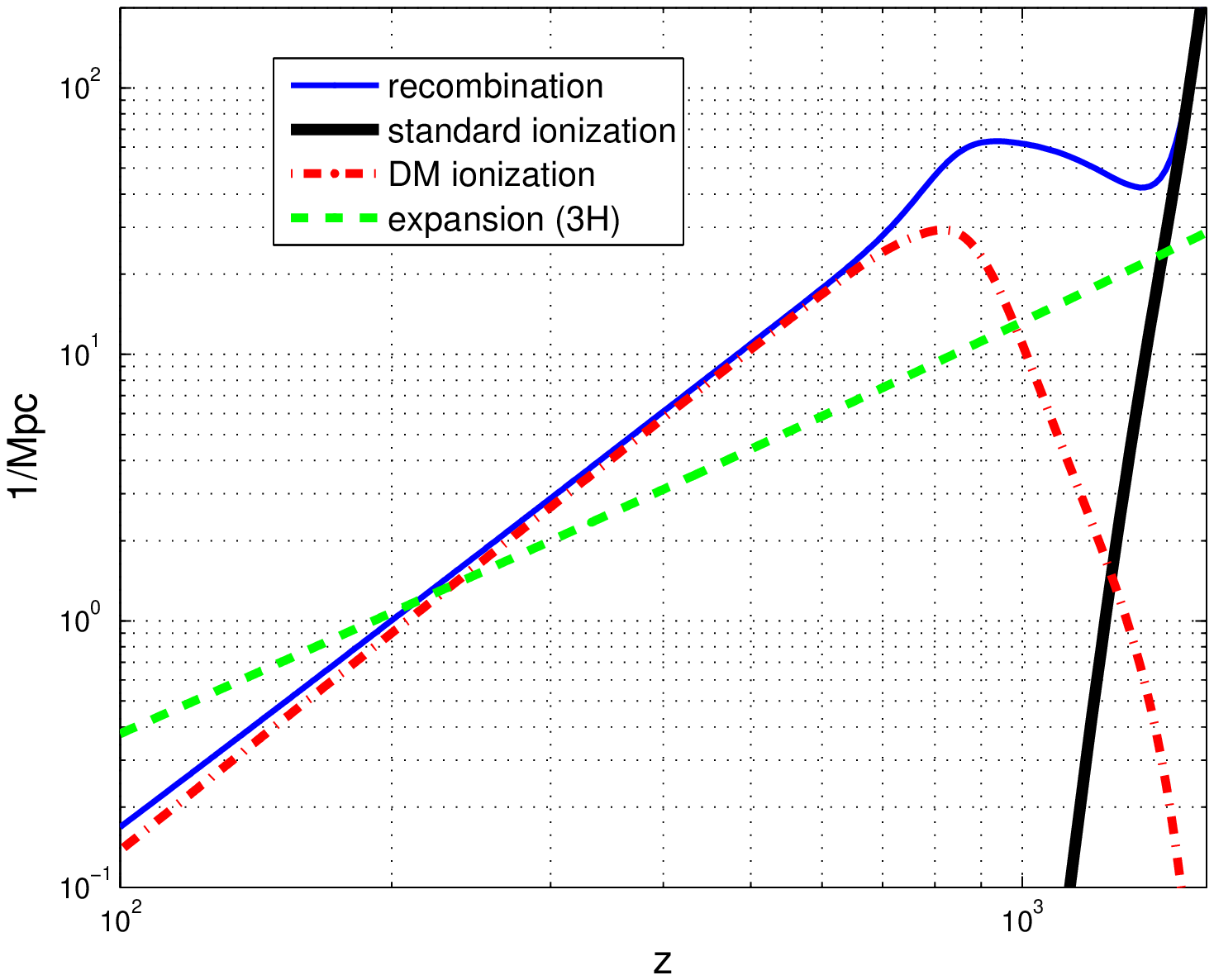}  \end{center}
\caption{Left: Ionization fraction, without DM annihilation (blue, solid; lowest curve) and with $\svm=1.5\times10^{-27}$~cm$^3$/s/GeV (green, solid; middle curve) and $\svm=1.5\times10^{-26}$~cm$^3$/s/GeV (red, solid; uppermost curve). We omit late time reionization in the plot. Dashed lines denote the floor approximation, Eq.~(\ref{eq:flr}). Right: Relevant time scales; recombination rate per electron $\alpha_Hn_e$ (blue solid thin curve), standard photoionization rate per electron $\beta_He^{-\epsilon_{12}/T_R}(x_e^{-1}-1)$ (black solid thick curve), DM ionization rate per electron $I_\chi/n_e$ (red dot-dashed), and expansion $3H$ (green dashed). We take $\svm=1.5\times10^{-26}$~cm$^3$/s/GeV.}
\label{fig:xet}
\end{figure}

Sample numerical solutions of Eq.~(\ref{eq:3lev}) are depicted by solid lines in the left panel of Fig.~\ref{fig:xet}, where we plot the ionization fraction,
\be x_e=\frac{n_e}{n_H},\ee
vs. redshift. 
As is well known, DM of mass $m_\chi=\mathcal{O}({\rm GeV})$ and annihilation cross section compatible with thermal freeze-out can have a significant effect on the ionization fraction after recombination. 

To understand the relevant processes, it is instructive to inspect the time scales appearing in Eq.~(\ref{eq:3lev}). In the right panel of Fig.~\ref{fig:xet} we consider DM with $\svm=1.5\times10^{-26}$~cm$^3$/s/GeV. We represent the DM ionization rate per electron by $(I_\chi/n_e)$, the standard ionization term (due to CMB photons) by $\beta_He^{-\epsilon_{12}/T_R}(x_e^{-1}-1)$, and recombination by $\alpha_Hn_e$. 
Notably, beginning at $z\sim1000$ and down to $z\sim200$, both DM ionization and recombination are faster than Hubble expansion.  
Thus, for large enough annihilation rate, $n_e$ follows a quasi-equilibrium solution balancing recombination off DM ionization alone,
\be\label{eq:flr} x_e^{\rm floor}=\frac{\rho_\chi}{\rho_b}\sqrt{\frac{16}{27}\frac{m_H^2}{m_\chi\epsilon_H}\frac{\langle\sigma v\rangle}{\alpha_H}},\ee
where we expanded to zeroth order in $x_e$. This is the ``floor solution" pointed out in~\cite{Padmanabhan:2005es}. Because $\alpha_H\propto z^{-2/3}$ at the relevant redshift, a constant velocity weighted annihilation cross section results in 
\be\label{eq:flrscaling} x_e^{\rm floor}\approx4.2\times10^{-3}\left(\frac{z}{1000}\right)^{1/3}\left(\frac{\langle\sigma v\rangle}{3\times10^{-26}~{\rm cm^3/s}}\right)^{1/2}\left(\frac{m_\chi}{\rm GeV}\right)^{-1/2}.\ee
In the left panel of Fig.~\ref{fig:xet} we depict the floor solutions by dashed lines.  

We caution the reader that DM with mass $m_\chi\sim1$~GeV and thermal relic annihilation cross section, as depicted by the red curves in Fig.~\ref{fig:xet}, is already excluded by CMB constraints. The strongest constraints we are currently aware of combine WMAP7 and SPT data to derive the bound $\svm\lesssim1.6\times10^{-27}$~cm$^3$s$^{-1}$GeV$^{-1}$ at 95\%CL~\cite{Giesen:2012rp}. WMAP7+ACT gives $\svm\lesssim2.1\times10^{-27}$~cm$^3$s$^{-1}$GeV$^{-1}$ at 95\%CL~\cite{Galli:2011rz}. WMAP7 alone gives $\svm\lesssim5.7\times10^{-27}$~cm$^3$s$^{-1}$GeV$^{-1}$ at 95\%CL~\cite{Hutsi:2011vx}. We thus use the example in Fig.~\ref{fig:xet} merely to highlight the physics. 

The important point to take home is that DM annihilation can easily dominate the fractional ionization immediately after recombination. This remains true as long as the DM-induced ionization rate is comparable to the expansion right after recombination, namely, as long as $\svm\gtrsim10^{-27}~{\rm cm^3s^{-1}GeV^{-1}}$, corresponding roughly to the green curve in Fig.~\ref{fig:xet}. Somewhat surprisingly, the current CMB constraints do allow for sufficient annihilation power. This is because most of the effect on the CMB temperature power spectrum on small angular scales, $l\gtrsim100$, is contained by an overall suppression factor $C_l\to e^{-2\Delta\tau}\,C_l$, where $\Delta\tau$ denotes the excess optical depth due to the extra ionization. This overall factor is degenerate with adjusting the amplitude of the primordial curvature power spectrum, $A_s\to e^{2\Delta\tau}\,A_s$~\cite{Padmanabhan:2005es}. The amplitude degeneracy is not complete, and is ameliorated by polarization data; nevertheless, additional degeneracy with the primordial tilt $n_s$ and with other cosmological parameters leads to the fact that CMB constraints still allow a much larger role for DM annihilation at recombination than could naively be guessed. 

The main simple result of this paper can be understood directly from Eq.~(\ref{eq:flr}). Generalizing to include cosmological perturbations, Eq.~(\ref{eq:flr}) tells us that in the quasi-equilibrium limit, we may expect the electron density perturbations to track DM perturbations,
\be\label{eq:simple}\delta_e=\frac{\delta n_e}{n_e}\sim\delta_\chi,\ee
where the DM density contrast is given by $\delta_\chi=(\delta\rho_\chi/\rho_\chi)$. During and soon after recombination, DM perturbations on small scales are orders of magnitude larger than the corresponding baryon perturbations, $\delta_\chi\gg\delta_b$, because the latter are trapped in acoustic oscillations until the end of the baryon drag epoch while the former simply collapse gravitationally since horizon entry. Thus we may expect a large enhancement in the free electron density $\delta_e$ compared to the Standard Model prediction. Here we neglected several factors, including e.g. photon and kinetic matter temperature perturbations. However, in the next section we show that the simple reasoning behind Eq.~(\ref{eq:simple}), motivated by the floor solution Eq.~(\ref{eq:flr}), is essentially correct.

Finally, DM annihilation affects also the kinetic matter temperature, though around the time of recombination this effect is much less important than the effect on the ionized fraction. The equation for matter temperature is given by\footnote{We omit negligible corrections associated with photorecombination/ionization cooling/heating~\cite{Switzer:2007sn}. We thank Yacine Ali-Ha\"{i}moud for a discussion on this point.}
\be\label{eq:mt}
{dT_M\over dt}+2HT_M\approx\frac{\gamma}{n_b+n_e},\;\;\;\;\;\gamma=\gamma_C+\gamma_\chi,
\ee
with
\ba\gamma_C&=&\frac{8\sigma_Ta_R}{3m_ec}\,n_eT_R^4(T_R-T_M),\\
\gamma_\chi&=&\frac{2}{3k_B}\frac{\langle\sigma v\rangle \rho_\chi^2}{m_\chi}\left({n_H+2n_e\over 3n_H}\right),\ea
where $a_R$ is the radiation constant.  

At redshifts $z\gtrsim200$, Compton scattering dominates and the matter temperature $T_M$ tracks the CMB temperature $T_R$. At those early times, DM annihilation has negligible effect on the matter temperature. This is different than what we have just seen for the ionized fraction, and will carry over to the perturbation analysis in the next section. The reason is that, compared to the power available from DM annihilation, the CMB energy reservoir is intense but cool. DM annihilation can dominate ionization, because ionization can only feed off the deep Boltzmann tail of the CMB spectrum. The matter kinetic energy, however, is driven by Thomson scattering off the bulk of the CMB spectrum and thus DM looses this battle by a large margin. Quantitatively, we readily see this by inspecting the ratio of heating rates,
\ba\frac{\gamma_C}{\gamma_\chi}&\approx&12\frac{\sigma_Tc}{\langle\sigma v\rangle}\frac{m_\chi T_R}{m_Hm_e}\frac{\rho_H\rho_\gamma}{\rho_\chi^2}\,x_e\left(1-\frac{T_M}{T_R}\right)\no\\
&\sim&2\cdot10^{4}\left(\frac{\langle\sigma v\rangle}{3\cdot10^{-26}{\rm cm^3/c}}\right)^{-1}\left(\frac{m_\chi}{\rm GeV}\right)\left(\frac{z}{10^3}\right)^2\left(\frac{x_e}{10^{-1}}\right)\left(1-\frac{T_M}{T_R}\right).\ea
This comparison means that around recombination, when $z\sim10^3$ and $x_e\sim10^{-1}-10^{-2}$, DM annihilation cannot break the relation $T_M=T_R$, enforced by Thomson scattering. Only later at $z\lesssim10^2$ and with $x_e\sim10^{-3}-10^{-4}$, can DM annihilation compete with CMB heating.  
At this later time baryons kinetically decouple from the CMB and DM annihilation can change the evolution of $T_M$ appreciably.

\subsection{Cosmological perturbations: inhomogeneous recombination}\label{ssec:inhomde}

We now compute the first order perturbations to the free electron density. Our aim is to refine the rough analysis leading to Eq.~(\ref{eq:simple}). As we discuss in more detail in Sec. \ref{sec: sec_order_anisotropies}, what motivates us in pursuing this calculation, is that electron density perturbations during recombination induce apparent non-gaussianity in the CMB anisotropies as measured by a late time observer~\cite{Senatore:2008vi,Senatore:2008wk,Khatri:2008kb,Khatri:2009ja}. This non-gaussianity can in principle be measurable by Planck and future experiments, offering a potential means to test our scenario. 

For simplicity we assume that the energy from DM annihilation is instantaneously absorbed by the plasma. As mentioned earlier, this can be a poor approximation, the extent to which it applies depending on model details. The smearing of energy absorption by the plasma leads to damping of small scale power. We analyze this issue in some detail in App.~\ref{app:deposition}.

We work in synchronous gauge,
\be ds^2=-a^2\left(d\eta^2-\left(\delta_{ij}+h_{ij}\right)dx^idx^j\right),\ee
fixing the gauge as usual by eliminating the DM velocity perturbations. Considering only scalar perturbations, we denote the trace and the trace-less parts of the scalar mode of $h_{ij}$ by $h$ and $\kappa$. (These correspond to $h$ and $\eta$ in the notation of Ma and Bertchinger~\cite{Ma:1995ey}.) 
Our normalization for the primordial curvature perturbation is such that $\xi_{\vec k}=1$ on superhorizon scales. 

The Boltzmann and Einstein equations for metric ($h,\kappa$), radiation ($\delta_{T_R}$), dark matter ($\delta_\chi$) and baryon density and velocity perturbations ($\delta_b$ and $v_b$), are not coupled to the electron and matter kinetic temperature perturbations ($\delta_e$ and $\delta_{T_M}$) at first order. 
Therefore, for all perturbations other than $\delta_e$ and $\delta_{T_M}$, we may use the usual set of Boltzmann and Einstein equations, given e.g. in Ma and Bertchinger~\cite{Ma:1995ey}. 

Given the solutions for $h,\;\kappa,\;\delta_{T_R},\;\delta_\chi,\;\delta_b,$ and $v_b$ -- amounting to the usual transfer functions -- we use them as sources for the linearized electron and matter temperature perturbations. Starting with the results of~\cite{Senatore:2008vi}, we add DM annihilation to obtain:
\ba
\dot\delta_e&=&\dot\delta_b+\frac{aQ_e}{n_e}\left(\sum_X\left(\frac{\partial\log Q_e}{\partial\log X}\right)\delta_X-\delta_e\right),\label{eq:de}\\
\dot\delta_{T_M}&=&-\frac{\dot h}{3}-\frac{2ik}{3}v_b+\frac{a\gamma}{(n_b+n_e)T_M}\left(\sum_X\left(\frac{\partial\log\gamma}{\partial\log X}\right)\delta_X-\delta_{T_M}-\frac{n_e\delta_e+n_b\delta_b}{n_b+n_e}\right).\label{eq:dTm}
\ea
Here $X=\left\{H,n_e,n_b,n_\chi,T_M,T_R\right\}$, and\footnote{Note that the $H-$dependence of $Q_e$ is contained in the $C_H$ factor for $L\alpha$ escape, Eq.~(\ref{eq:CH}). For the matter temperature, we have $(\partial\log\gamma/\partial\log H)=0$.} $\delta_H\equiv-\left(\dot\delta_b/3aH\right)$ denotes the perturbation to the baryon velocity divergence, as measured by a comoving local observer. 

It is straightforward to solve Eqs.~(\ref{eq:de}-\ref{eq:dTm}) numerically. However, a simplification occurs if one suffices with computing observable effects in the CMB\footnote{The reasoning here may well need to be modified if one aims to address physics at later epochs, e.g. for $21$ cm analyses. For such late time effects one needs to solve Eqs.~(\ref{eq:de}) and~(\ref{eq:dTm}) simultaneously -- as we do were required in this paper.}. As discussed in the previous section, the matter temperature is clipped to the radiation temperature around last scattering. This partially carries over to the perturbations: $\delta_{T_M}\approx\delta_{T_R}$, all the way until the end of the baryon drag epoch when $\delta_{T_M}$ rises by compression as the baryons fall into the DM potential wells. By the time $\delta_{T_M}$ finally breaks loose of $\delta_{T_R}$, then, the Thomson optical depth for photons is small and the electron perturbation has little residual effect on the observed CMB anisotropy. 

Setting $\delta_{T_M}=\delta_{T_R}$, we can write a direct integral solution for $\delta_e$,
\ba
\label{eq:dif}
\delta_e(k,\eta)&=&\int_{\eta_{\rm init}}^\eta d\eta'\mathcal{G}_e(k,\eta')\exp\left(-\int_{\eta'}^{\eta}d\eta''\mathcal{F}_e(\eta'')\right),\\
\mathcal{F}_e&=&\frac{aQ_e}{n_e}\left(1-\left(\frac{\partial\log Q_e}{\partial\log n_e}\right)\right),\;\;\;\;
\mathcal{G}_e=\dot\delta_b+\frac{aQ_e}{n_e}\,\sum_{X'}\left(\frac{\partial\log Q_e}{\partial\log X}\right)\delta_{X'},\no
\ea
where the sum over $X'$ now does not include $n_e$. The initial time $\eta_{\rm init}$ is chosen early enough so that $\delta_e(k,\eta_{\rm init})=\delta_b(k,\eta_{\rm init})$. Eq.~(\ref{eq:dif}) allows us to obtain $\delta_e$ directly and quickly, reading all other perturbations from the numerical code CAMB~\cite{camb_notes,Lewis:1999bs,Howlett:2012mh}; it agrees well with the full numerical solution to Eq.~(\ref{eq:de}) throughout and for a good while after recombination. 

As expected from the discussion in the previous section, the calculations confirm the presence of a growing ionization mode sitting on top of the usual baryon acoustic oscillations (BAOs). We now examine this result in some detail.

\begin{figure}[!t]\begin{center}
\includegraphics[width=0.475\textwidth]{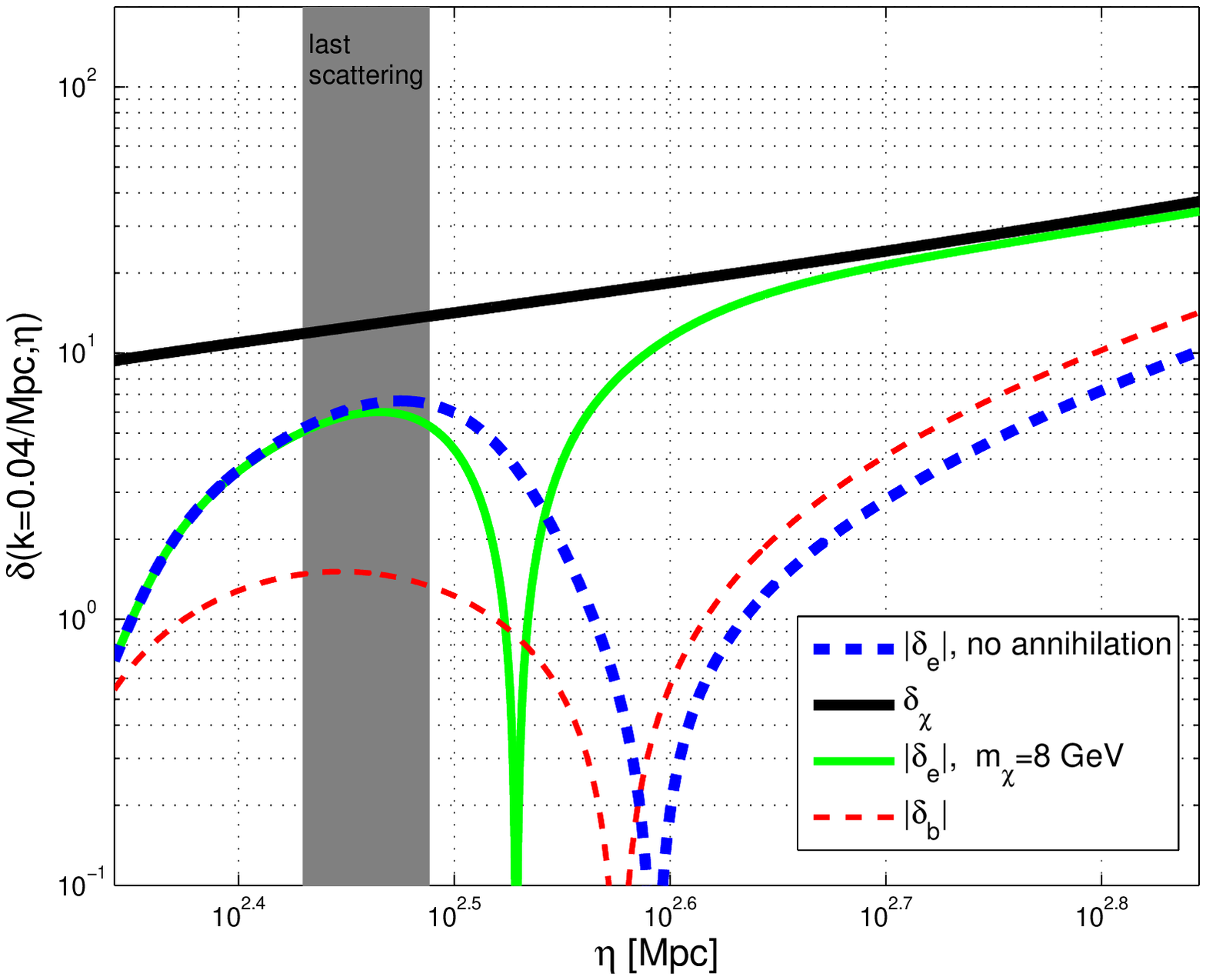}\quad
\includegraphics[width=0.475\textwidth]{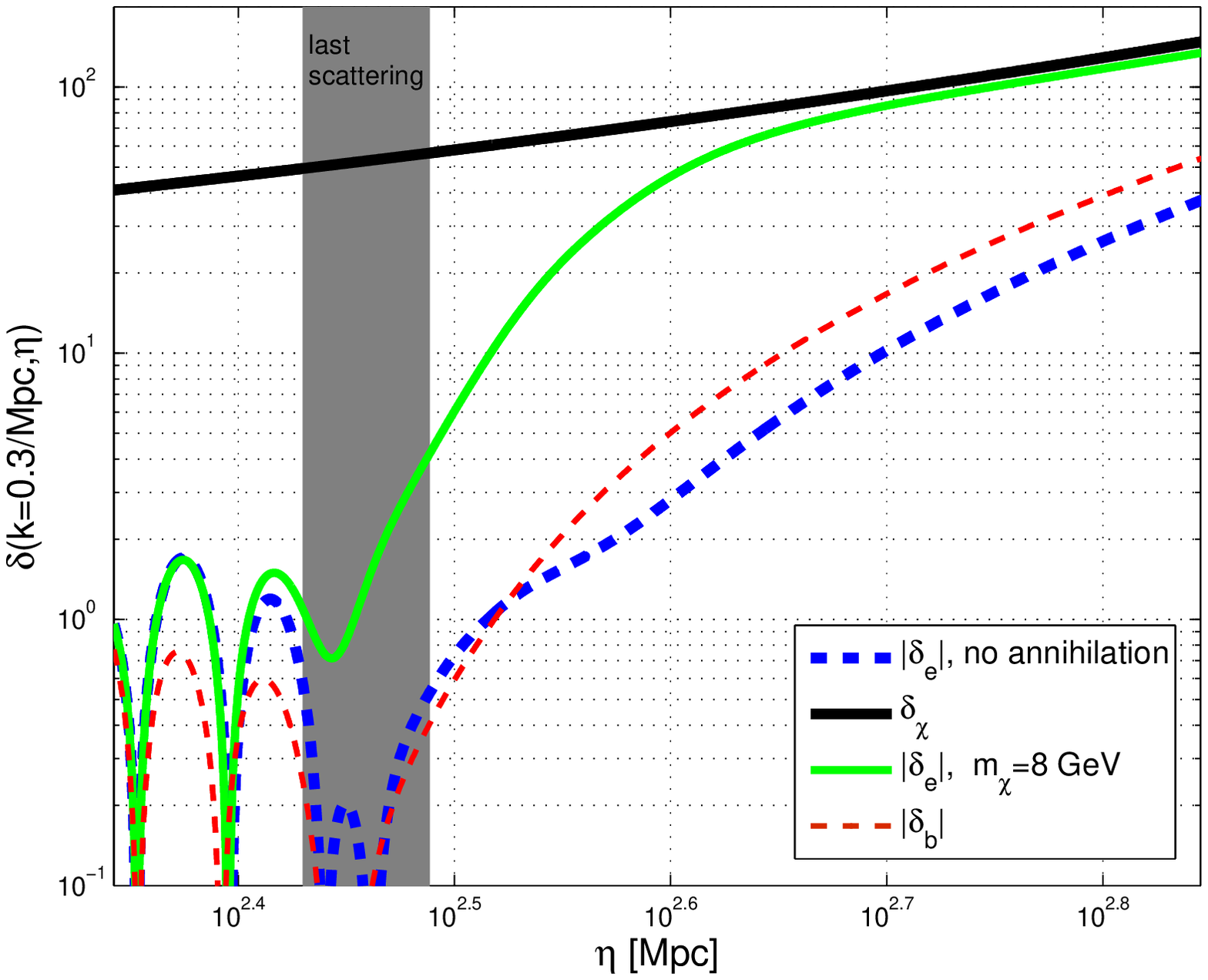}  \end{center}
\caption{Perturbations vs. conformal time. The curves are: $\delta_\chi$ (solid thick black); $|\delta_b|$ (dashed thin red);  $|\delta_e|$ with $m_\chi=8$~GeV (solid green); $|\delta_e|$ with no annihilation (dashed thick blue). For the $m_\chi=8$~GeV curve, thermal relic annihilation cross section is assumed. Grey shaded band denotes full width half-maximum of the visibility function. Left: wave number $k=0.04$~Mpc, corresponding to $l\sim600$. Right: $k=0.3$~Mpc, corresponding to $l\sim4.2\times10^3$. Synchronous gauge; $\xi=1$ on superhorizon scales.}
\label{fig:pertt}
\end{figure}

We begin with an eye towards observability in the CMB. In Fig.~\ref{fig:pertt} we fix the wave number of the perturbation and examine its time dependence. Green line shows the electron perturbation with $\svm=3.75\times10^{-27}$~cm$^3$s$^{-1}$GeV$^{-1}$; blue shows the electron perturbation with no DM annihilation. For reference, we show also the baryon and DM density perturbations in grey and black, respectively. Note that for $\delta_e$ and $\delta_b$, we plot the absolute value of the transfer functions, while the DM transfer function is positive.
The grey shaded band depicts the full width half-maximum (FWHM) of the visibility function (taken here with no DM annihilation). In the left panel we fix $k=0.04$~Mpc$^{-1}$, corresponding to observed anisotropy multipole $l\sim k\eta_0\sim600$, where $\eta_0\sim1.4\times10^4$~Mpc is the conformal time today. In the right panel we fix $k=0.3$~Mpc$^{-1}$, corresponding to $l\sim4200$. 

The growing ionization mode due to DM annihilation is clearly visible. This mode grows towards, and finally catches up with the DM perturbations, eventually  amplifying $\delta_e$ by more than an order of magnitude compared with the Standard Model prediction. However, it takes finite time for the quasi-equilibrium configuration to manifest itself, particularly so on larger scales; this causes much of the amplification effect of $\delta_e$ to only take place after CMB last scattering. In addition, note that dragging the electron perturbation in the positive direction towards the DM perturbation, can actually lead to suppression of the magnitude -- in absolute value -- for perturbations that enter the last scattering surface with negative amplitude. This is seen in the left panel of Fig.~\ref{fig:pertt}.  
%

\begin{figure}[!t]\begin{center}
\includegraphics[width=0.475\textwidth]{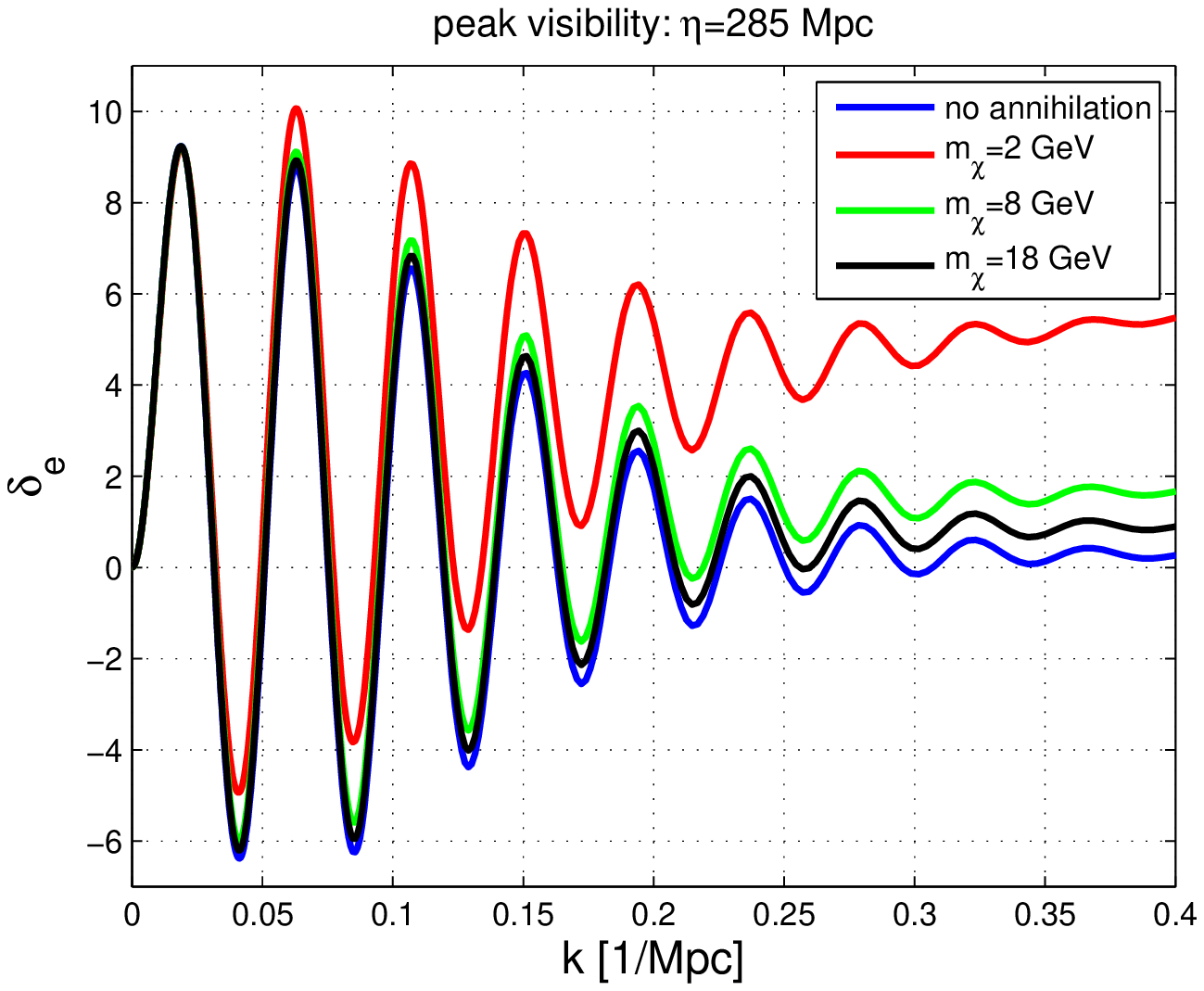}\quad
\includegraphics[width=0.475\textwidth]{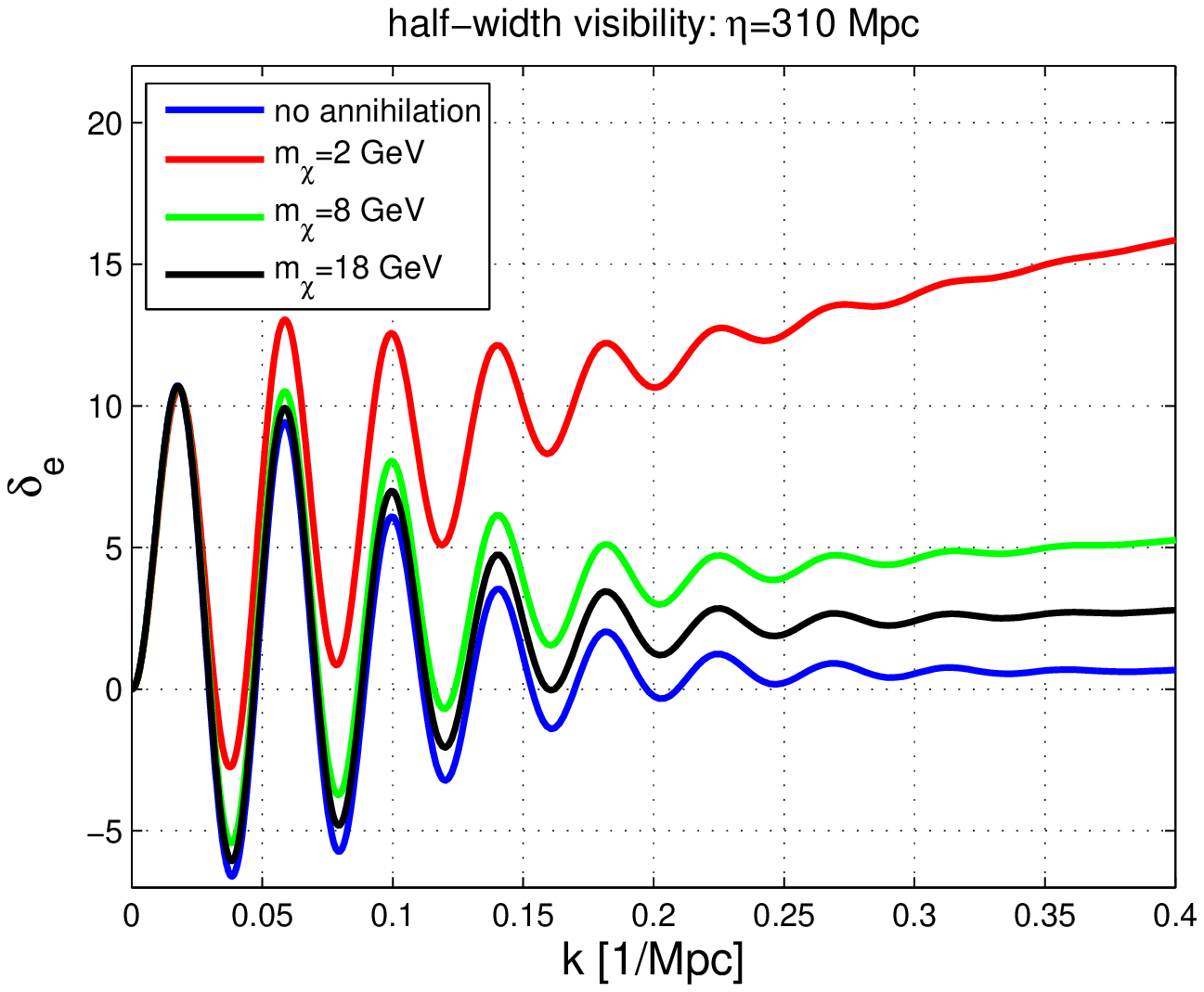}  \end{center}
\caption{Electron perturbation vs. wave number, at peak (left) and half-maximum (right) of visibility. Thermal relic annihilation cross section is assumed. The curves are, from top to bottom: $m_\chi=2,8,18$~GeV, and no annihilation. Synchronous gauge; $\xi=1$ on superhorizon scales. We caution the reader that the red curve with $m_\chi=2$~GeV is excluded experimentally by CMB data, and is only shown here for illustration.}
\label{fig:pertrec}
\end{figure}

In Fig.~\ref{fig:pertrec} we study the scale dependence. In the left panel, we show the $k$ dependence of the perturbations in a snapshot close to the peak of visibility $\eta=285$~Mpc. On the right we focus on the half-maximum width, $\eta\sim310$~Mpc. The effect is larger for larger wavenumber, as DM perturbations on smaller scales enter the horizon earlier and have more time to collapse before recombination, leading to more efficient ionization. In addition, as noted above, the growing mode becomes significant only somewhat after the peak of the visibility. We caution the reader again that the red curve with $m_\chi=2$~GeV is excluded experimentally by CMB data, and is only shown here for illustration.

\begin{figure}[!t]\begin{center}
\includegraphics[width=0.475\textwidth]{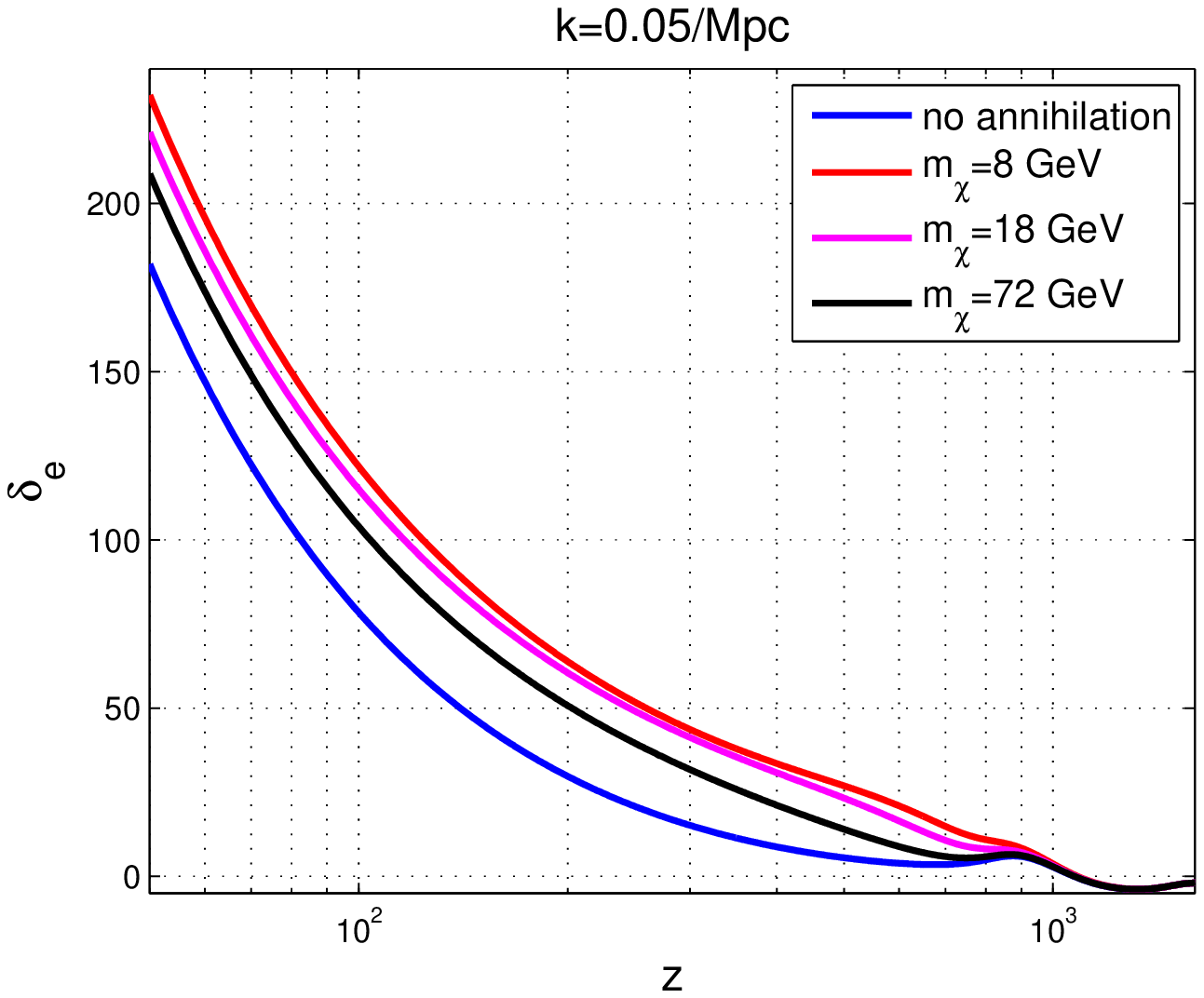}\quad
\includegraphics[width=0.475\textwidth]{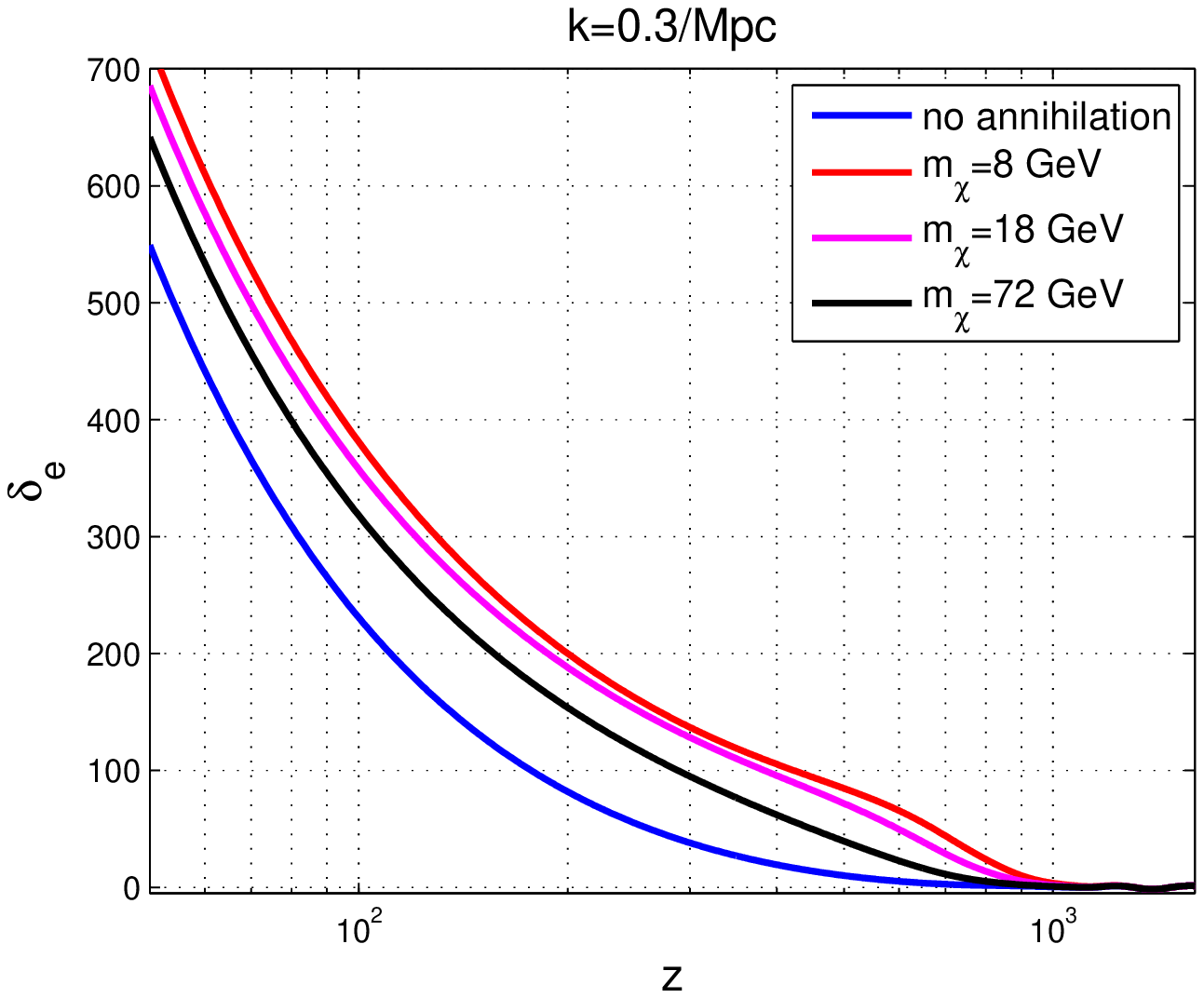}  \end{center}
\caption{Electron perturbations vs. redshift. Left: wavenumber $k=0.05$~Mpc, corresponding to $l\sim700$. Right: $k=0.3$~Mpc, corresponding to $l\sim4.2\times10^3$. Thermal relic annihilation cross section is assumed. The curves are, from top to bottom: $m_\chi=8,18, 72$~GeV, and no annihilation. Synchronous gauge; $\xi=1$ on superhorizon scales.}
\label{fig:pertz}
\end{figure}

In Fig.~\ref{fig:pertz} we show $\delta_e$ vs. redhsift at times significantly after CMB last scattering, again for two different wavenumbers. Fig.~\ref{fig:pertz} concerns the deep Dark Ages; the detailed dynamics at last scattering, crucial for CMB analyses, is merely seen as small wriggles around $z\sim1100$. Here we confirm the naive estimate of Eq.~(\ref{eq:simple}), that says that for high enough annihilation power, the electron perturbation reaches quasi-equilibrium and roughly sticks to the DM perturbation. In this regime, $\delta_e$ is roughly independent of the DM mass and annihilation rate and is boosted by a factor 2-3 compared with the Standard Model prediction. Reducing the annihilation power below $\svm\sim10^{-27}$~cm$^{3}$s$^{-1}$GeV$^{-1}$, as seen for $m_\chi=36$~GeV, causes ionization and recombination to drop below the Hubble rate, freezing the ionization fraction below the quasi-equilibrium value. 

\begin{figure}[!t]\begin{center}
\includegraphics[width=0.475\textwidth]{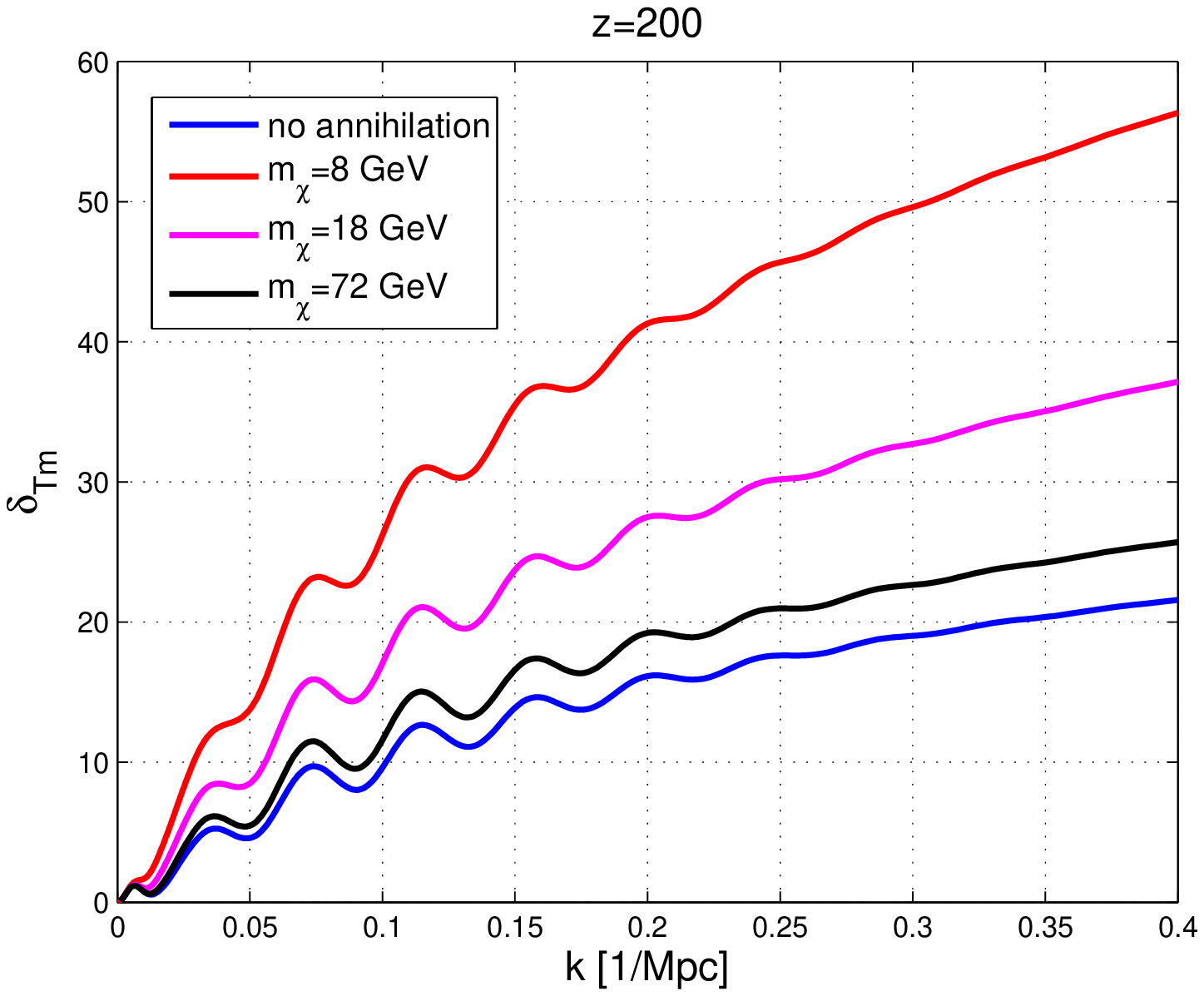}\quad
\includegraphics[width=0.475\textwidth]{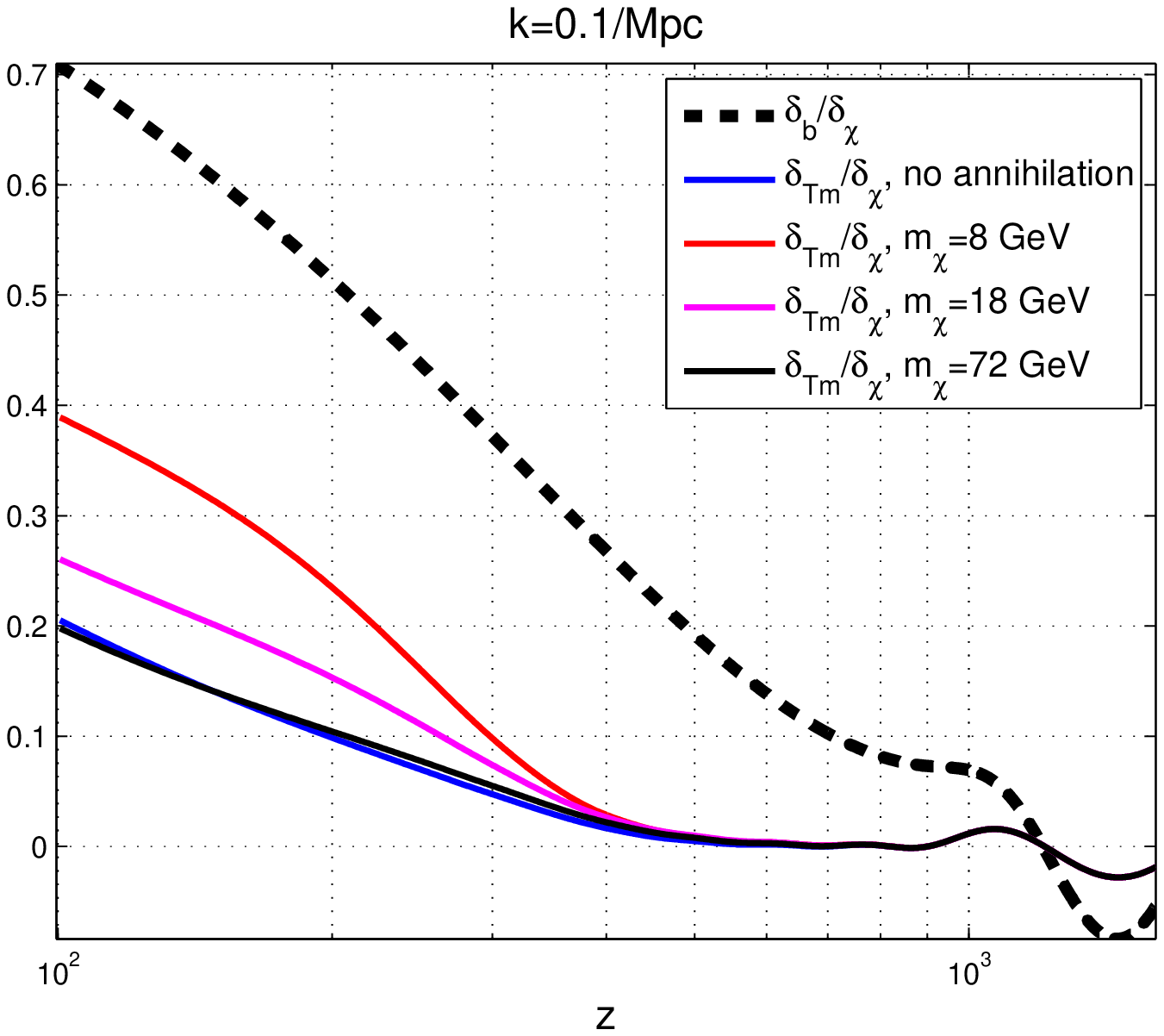}  \end{center}
\caption{Left: matter temperature perturbation vs. wavenumber. Thermal relic annihilation cross section is assumed. The curves are, from top to bottom: $m_\chi=8,18, 72$~GeV, and no annihilation. Right: baryon and temperature perturbations relative to dark matter perturbation. The top (dashed) curve shows $\delta_b/\delta_\chi$. The bottom four curves are as in the left panel. Synchronous gauge; $\xi=1$ on superhorizon scales.}
\label{fig:21cm}
\end{figure}

During the dark ages, the relevant future probe of DM annihilation would be in the absorption of $21$ cm radiation~\cite{Furlanetto:2006wp,Valdes:2007cu,Cumberbatch:2008rh,Finkbeiner:2008gw,Natarajan:2009bm,Valdes:2012zv}.  (See Ref.~\cite{Furlanetto:2006jb} for a review.) Here, the relevant quantity is the matter temperature entering the computation of the spin temperature~\cite{Loeb:2003ya,Lewis:2007kz}. In the left panel of Fig.~\ref{fig:21cm} we plot the matter temperature perturbation at $z=200$ as a function of wavenumber. We learn that a factor of $\sim2$ enhancement in $\delta_{T_M}$ can arise from DM annihilation. In the right panel, we plot the ratio of the baryon and matter temperature perturbations to the DM perturbation as a function of redshift, for fixed wavenumber $k=0.1$~Mpc$^{-1}$. As an aside, using Eqs.~(\ref{eq:de}-\ref{eq:dTm}) we can solve for the matter temperature perturbations from the correct initial conditions at recombination down to the deep dark ages. Doing this, we note that the scale-independent relation $\delta_{T_M}(k,\eta)=s(\eta)\,\delta_{b}(k,\eta)$ with $s(\eta)$ independent of $k$, assumed e.g. in~\cite{Furlanetto:2006wp} and later references, is violated at $\mathcal{O}(1)$. 

Previous analyses of CMB non-gaussianity induced by perturbations to the free electron density around recombination~\cite{Senatore:2008vi,Senatore:2008wk,Khatri:2008kb,Khatri:2009ja}, have found a level of non-gaussianity that could be marginally detectable by Planck. We have seen that DM annihilation could boost small scale electron perturbations by a sizable amount. It is thus interesting to assess the bispectrum when DM annihilation is taken into account. 

As we show in Sec.~\ref{sec: sec_order_anisotropies}, despite the amplified electron perturbations, the DM annihilation effect on the CMB bispectrum is small. This comes about from three unfortunate reasons. First, photon diffusion acts to erase power on small angular scales, where the DM effect is pronounced. Second, as we saw recombination has a finite response time. By the time the DM-induced amplification reaches its full swing, photon last scattering is mostly over. And third, Thomson scattering cannot transmit power from a short wave electron perturbation down to a long wave photon anisotropy\footnote{In the relevant limit of photon number conservation.}. Thus, short wave electron perturbations do not contribute directly to the bispectrum in the squeezed limit, where much of the signal-to-noise is contained. At the end of the day, our results indicate that the large boost to $\delta_e$ will be very difficult to detect in the CMB even if the recombination bispectrum is measured, at least in the foreseeable future.

\section{Non Gaussianity: can we observe enhanced small scale electron perturbations in the CMB?}
\label{sec: sec_order_anisotropies}

Assuming gaussian initial conditions, any observed non-gaussianity and, in particular, a finite bispectrum comes about at second order in perturbation theory. In this section we write approximate formulae for the second order temperature anisotropies in the presence of (first order) electron density perturbations, and estimate the induced bispectrum. 

We stress that the analysis presented here is meant as a rough estimate of the observability of the effect highlighted in the previous section. We defer a more comprehensive analysis to~\cite{BDZ}. There, special care is given to electron perturbations on small scales, that have not been accounted for by existing analytical studies. The results then apply generically and no special treatment is required to include DM annihilation, beyond utilizing the modified $\delta_e$ computed above. 

To save the reader from disappointment: our final answer to the question posed in the title of this section is negative. We find that even an $\mathcal{O}(10)$ enhancement, compared to the Standard Model, in small scale electron perturbations, will leave only a very subtle imprint on the bispectrum. As most of the DM annihilation effect during last scattering is concentrated on such small scales, the CMB bispectrum will likely not provide means of detection.

\subsection{The bispectrum}

Our notation for the homogeneous (unperturbed) differential and integrated optical depth and the visibility function are given by
\be\dot\tau(\eta)=-ac\sigma_Tn_e,\;\;\;\;\tau(\eta)=-\int_\eta^{\eta_0}d\eta'\dot\tau(\eta'),\;\;\;\;g(\eta)=-e^{-\tau(\eta)}\dot\tau(\eta).\ee
We write the Fourier space temperature anisotropy as \be\Theta(\vec k,\eta,\hat n)=\Theta^{(1)}(\vec k,\eta,\hat n)+\Theta^{(2)}(\vec k,\eta,\hat n).\ee 
We neglect second order metric perturbations. Then, the first and second order anisotropies today are given by the line of sight (LOS) solutions:
\ba\label{eq:boltz1}\Theta^{(1)}(\vec k,\eta_0,\hat n)&=&\int_0^{\eta_0}d\eta e^{ik\mu_k(\eta-\eta_0)}S^{(1)}(\vec k,\eta,\hat n),\\
\label{eq:boltz2}\Theta^{(2)}(\vec k,\eta_0,\hat n)&=&\int_0^{\eta_0}d\eta e^{ik\mu_k(\eta-\eta_0)}\,g(\eta)\,\left(S_{\delta g}(\vec k,\eta,\hat n)+S^{(2)}(\vec k,\eta,\hat n)\right),\ea
where we define $\mu_k=\hat k\cdot\hat n$. The source terms are:
\ba
&\!\!S^{(1)}(\vec k,\eta,\hat n)&=g\left(\Theta_0^{(1)}(\vec k)+\mu_kv_b^{(1)}(\vec k)-\frac{1}{2}\mathcal{P}_2(\mu_k)\Pi^{(1)}(\vec k)+2\dot\alpha(\vec k)\right)+e^{-\tau}\left(\dot\kappa(\vec k)+\ddot\alpha(\vec k)\right)+\dot g\alpha(\vec k),\;\;\;\;\;\;\;\label{eq:s1}\\
&\!\!S_{\delta g}(\vec k,\eta,\hat n)&=\int\frac{d^3q}{(2\pi)^3}\delta_e(\vec k-\vec q)\left(\Theta^{(1)}_0(\vec q)+\mu_qv_b^{(1)}(\vec q)-\frac{1}{2}\mathcal{P}_2(\mu_q)\Pi^{(1)}(\vec q)-\Theta^{(1)}(\vec q,\hat n)\right),\;\;\;\;\;\;\;\label{eq:ds}\\
&\!\!S^{(2)}(\vec k,\eta,\hat n)&=
\Theta_{0}^{(2)}(\vec k)+\hat n\cdot\vec v_b^{(2)}(\vec k)-\frac{1}{2}\mathcal{P}_2(\mu_k)\Pi^{(2)}(\vec k).\;\;\;\;\;\;\;\label{eq:s2}\ea
We define $\alpha=(\dot h+6\dot\kappa)/2k^2$~\cite{Zaldarriaga:1996xe} and suppress the $\eta$ dependence on the RHS for clarity. The first order baryon velocity perturbation is assumed to be irrotational, $\vec v_b^{(1)}(\vec k,\eta)=\hat k v_b^{(1)}(\vec k,\eta)$.  

In Eq.~(\ref{eq:s2}) we neglect vector and tensor contributions ($m=\pm1,\pm2$, respectively; note that the rotational velocity vanishes in the bispectrum). Then, in the source terms, we can use the same Legendre multipole decomposition for first and second order terms,
\ba\label{eq:leg}
\Theta_l(\vec k,\eta)&=&\frac{i^l}{4\pi}\int d\hat n\mathcal{P}_l(\hat n\cdot\hat k)\Theta(\vec k,\eta,\hat n),
\ea
where $\mathcal{P}_l(x)$ are Legendre polynomials and with similar decomposition for the polarization $\Theta_{Pl}$, feeding into $\Pi=\Theta_2+\Theta_{P0}+\Theta_{P2}$. 
Note that for the second order perturbation, $\Theta^{(2)}$, the Legendre decomposition of Eq.~(\ref{eq:leg}) contains only part of the information because at second order, azimuthal symmetry around the wave vector no longer holds. Thus when we compute spherical harmonic coefficients, we will need to use the full $Y_{lm}$ transform for $\Theta^{(2)}$ (see Eq.~(\ref{eq:alm2}) below). Nevertheless, focusing on scalar contributions, the Legendre moments in Eq.~(\ref{eq:s2}) suffice to compute the second order source $S^{(2)}$ because there azimuthal averaging occurs by Thomson scattering.

Physically, the contribution $S_{\delta g}$ comes about from perturbing $\dot\tau$ and $\tau$ in the first order solution, Eq.~(\ref{eq:boltz1}). This term is equivalent to perturbing the visibility function, up to corrections proportional to the integrated Sachs-Wolfe effect that are only relevant on large scales. As written in Eq.~(\ref{eq:ds}), this term is ready to deploy in an explicit calculation of the three-point function.

The contribution $S^{(2)}$ contains the second order feedback. Namely, it includes the actual effect of the electron perturbation on the photon field, rather than, as in the previous term, the effect of perturbing the way we see that field today. We defer a derivation of $S^{(2)}$ to subsequent work~\cite{BDZ}. Our results for this term extend and improve the analysis of~\cite{Senatore:2008wk}, and disagree with~\cite{Khatri:2009ja}. 

In what follows, we derive the bispectrum contribution due to the visibility term $S_{\delta g}$ and discuss the effect of DM annihilation. We then discuss qualitatively the contribution due to the second term $S^{(2)}$. 

The spherical harmonic coefficients $a_{lm}=a_{lm}^{(1)}+a_{lm}^{(2)}$ are given by
\ba\label{eq:alm1} a_{lm}^{(1)}&=&4\pi\int\frac{d^3k}{(2\pi)^3}(-i)^l\Theta^{(1)}_l(\vec k,\eta_0)Y^*_{lm}(\hat k),\\
\label{eq:alm2}a_{lm}^{(2)}&=&\int\frac{d^3k}{(2\pi)^3}\,\int d\hat nY_{lm}^*(\hat n)\,\Theta^{(2)}(\vec k,\hat n,\eta_0).\ea
Using Eqs.~(\ref{eq:alm1}-\ref{eq:alm2}) we compute the bispectrum, $B^{\ell_1\ell_2\ell_3}_{m_1m_2m_3}=\langle a_{\ell_{1} m_1}a_{\ell_{2} m_2}a_{\ell_{3} m_3}\rangle$. After the dust settles, we find that the contribution due to $S_{\delta g}$ leads to the following result\footnote{Our derivation of Eq.~(\ref{eq:full_bispectrum}) follows closely that of~\cite{Khatri:2008kb,Khatri:2009ja}, but the result disagrees with theirs in a few terms.}:
\ba\label{eq:full_bispectrum}
\!\!\!\!\!\!\!\!B^{\ell_1\ell_2\ell_3}_{m_1m_2m_3}&=&\mathcal{G}_{m_1m_2m_3}^{l_1l_2l_3}\,\times\,\frac{4}{\pi^2}\int_0^{\eta_0}d\eta g(\eta) \left(f_{\ell_1}(\eta)g_{\ell_2}(\eta)+{\rm five\;permutations}\right),\\
g_\ell(\eta)&=&\int dkk^2P(k)\Theta^{(1)}_\ell(k,\eta_0)\,\jl{\ell}{}{}\delta_e(k,\eta),\no\\
f_\ell(\eta)&=&(-1)^l\int dkk^2P(k)\Theta^{(1)}_\ell(k,\eta_0)\,\sum_{l',l''}(2l'+1)(2l''+1)\wj{\ell}{\ell'}{\ell''}{0}{0}{0}^2i^{l+l'+l''}j_{l'}[k(\eta_0-\eta)]\no\\&\times&\left(\delta_{l''1}\frac{\theta_b^{(1)}(k,\eta)-\theta_\gamma^{(1)}(k,\eta)}{3k}+\delta_{l''2}\frac{\Pi^{(1)}(k,\eta)}{10}-\left(1-\delta_{l''0}\right)\left(1-\delta_{l''1}\right)\Theta^{(1)}_{l''}(k,\eta)\right).\no
\ea
In Eq.~(\ref{eq:full_bispectrum}), by $\delta_e(k,\eta),\,\Theta_l(k,\eta),$ etc. we refer to transfer functions, namely, we mean to have mod out the random initial curvature perturbation $\xi_{\vec k}$ from $\delta_e(\vec k,\eta),\,\Theta_l(\vec k,\eta).$ The variables $\theta_b=ikv_b$ and $\theta_\gamma=3k\Theta_1$ are as in~\cite{Ma:1995ey}.  We assume gaussian adiabatic initial curvature perturbations, with power spectrum 
\be\langle\xi_{\vec k}\xi_{\vec p}\rangle=(2\pi)^3\delta^{(3)}\left(\vec k+\vec p\right)P(k).\ee
Finally, the gaunt coefficient is
\be\mathcal{G}_{m_1m_2m_3}^{l_1l_2l_3}=\sqrt{\frac{(2\ell_1+1)(2\ell_2+1)(2\ell_3+1)}{4\pi}}\wj{\ell_1}{\ell_2}{\ell_3}{0}{0}{0}\wj{\ell_1}{\ell_2}{\ell_3}{m_1}{m_2}{m_3}.\ee

\begin{figure}[!t]\begin{center}
\includegraphics[width=0.475\textwidth]{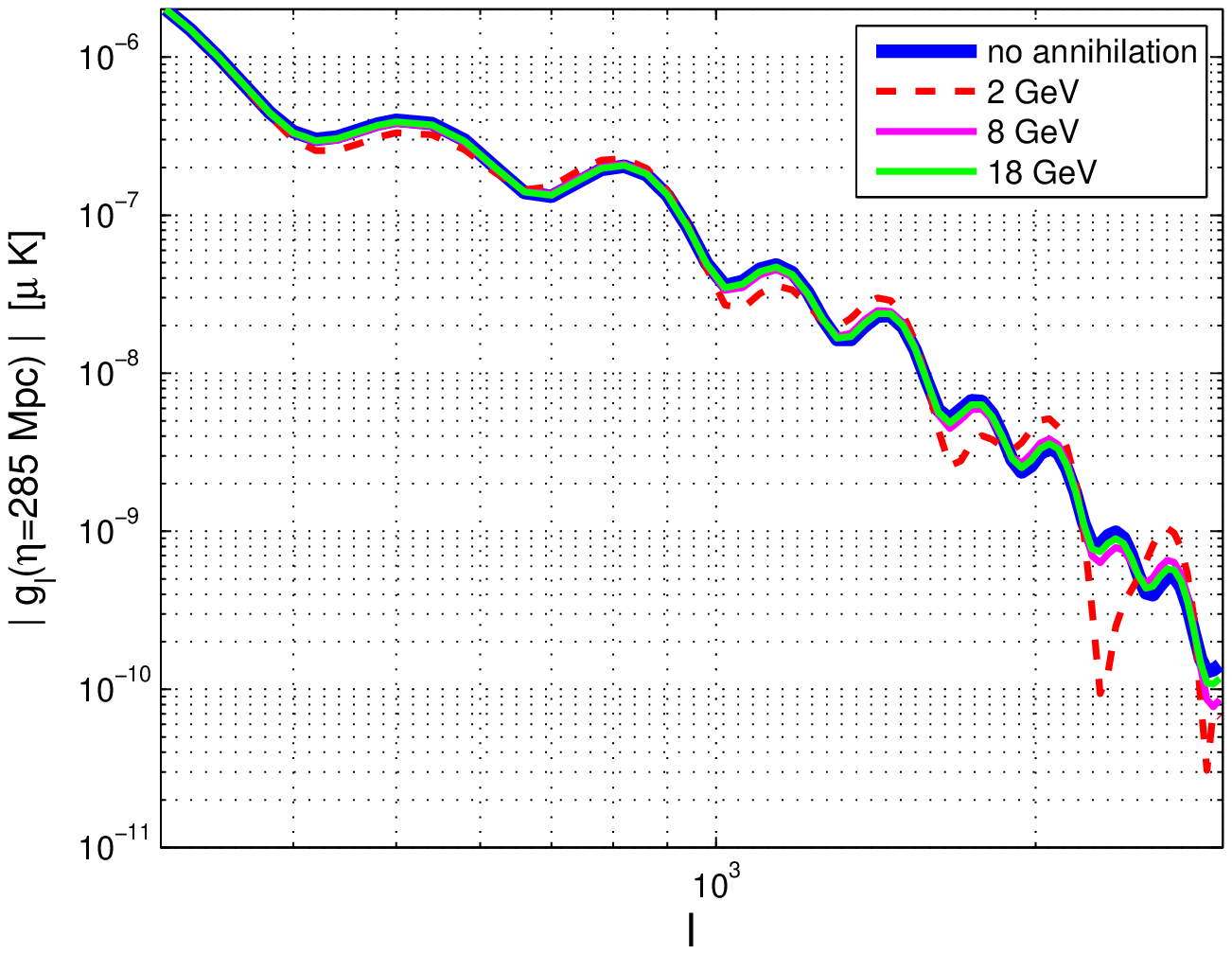}\quad
\includegraphics[width=0.475\textwidth]{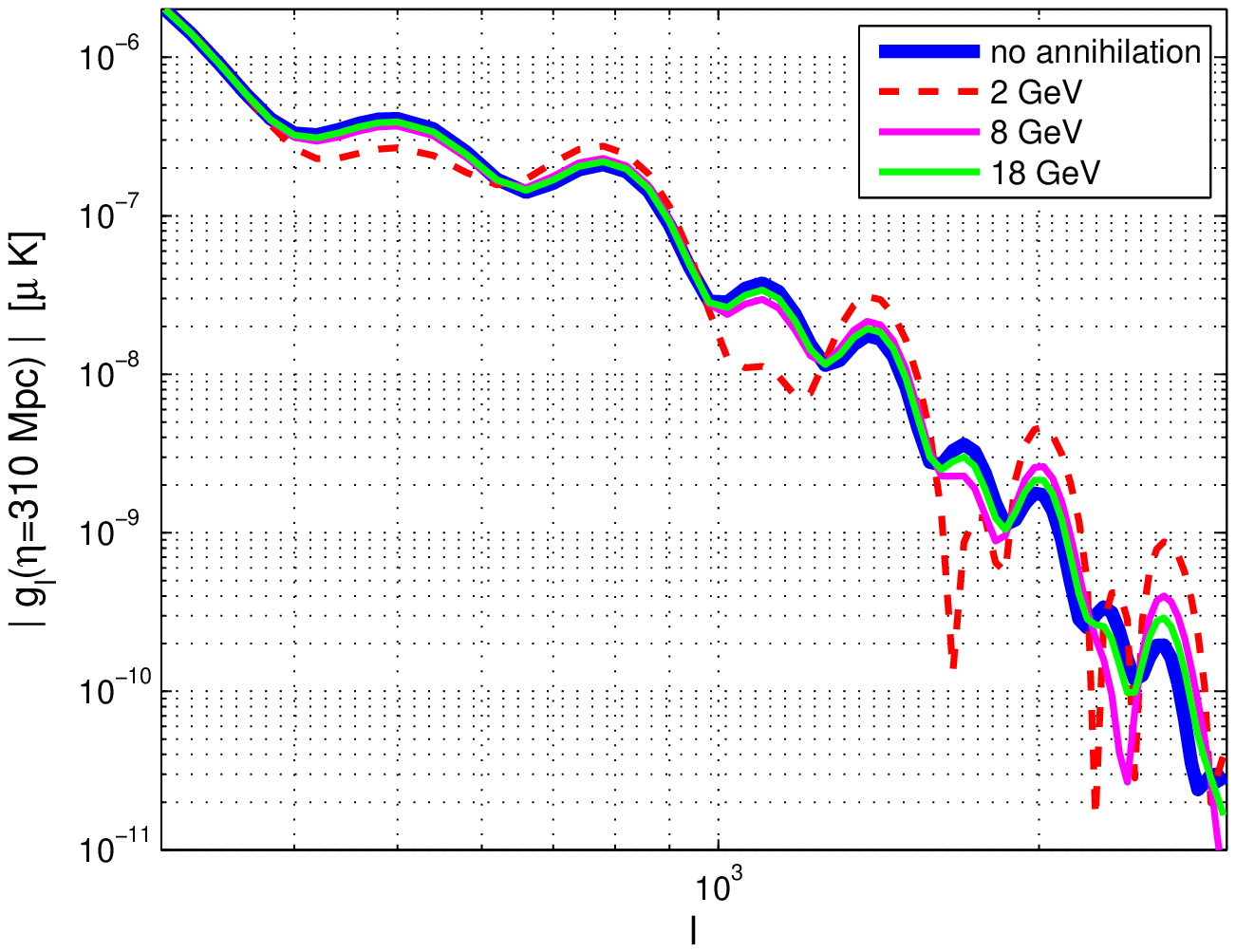}  \end{center}
\caption{The function $g_l(\eta)$ of Eq.~(\ref{eq:full_bispectrum}), capturing the DM annihilation effect in the visibility contribution to the recombination bispectrum. We plot $g_l$ (in absolute value, units of $\mu K$) vs. $l$, with a snapshot at peak visibility (left) and half-maximum visibility (right), for different DM masses assuming thermal freeze out cross section.}
\label{fig:grec}
\end{figure}

The information about the electron density perturbation is contained in the function $g_l(\eta)$. DM annihilation affects $g_l(\eta)$ at high $l$. In Fig.~\ref{fig:grec} we plot $g_l(\eta)$ (in absolute value, $\mu K$ units) vs. $l$, with a snapshot at peak visibility (left) and half-maximum visibility (right). We examine different DM masses, assuming standard thermal freezeout annihilation cross section. As it turns out, the growing ionization mode develops too late to show up significantly during last scattering where the visibility is large, and on angular scales that are too small to be accessible to current and upcoming experiments. Only in the $2$~GeV case, that is already excluded experimentally, a noticeable effect occurs at peak visibility and on scales $l\lesssim2000$. Imposing existing constraints, the effect is small even on scales $l\gtrsim2000$, beyond the reach of Planck. 

Obviously the effect of DM annihilation in the bispectrum is small, on scales $l\lesssim2000$ where other secondaries such as point sources are under control (see e.g.~\cite{Komatsu:2001rj}). Nevertheless, for completeness we estimate the signal-to-noise. 
The signal-to-noise ratio for the detection of the bispectrum, for an experiment with detector noise $N_\ell$ and covering a fraction of the sky $f_{\rm sky}$, is given by
\be
\left(\frac{S}{N}\right)^2=\sum_{\ell_{min}\leq \ell_1\leq \ell_2\leq \ell_3\leq \ell_{max}}\frac{f_{\rm sky}\,\left(B^{\ell_1\ell_2\ell_3}\right)^2}{\Delta_{\ell_1\ell_2\ell_3}\left(C_{\ell_1}+N_{\ell_1}\right)\left(C_{\ell_2}+N_{\ell_2}\right)\left(C_{\ell_3}+N_{\ell_3}\right)},
\ee
with $\Delta_{\ell_1\ell_2\ell_3}=1,2,6$ for zero, two and three equal $\ell$'s. The reduced bispectrum is obtained by summing over $m$-modes,
\be B^{\ell_1\ell_2\ell_3}=\sum_{m_1,m_2,m_3}\wj{\ell_1}{\ell_2}{\ell_3}{m_1}{m_2}{m_3}B^{\ell_1\ell_2\ell_3}_{m_1m_2m_3}.\ee

\begin{figure}[!ht]
\centering
\includegraphics[width=0.5\textwidth]{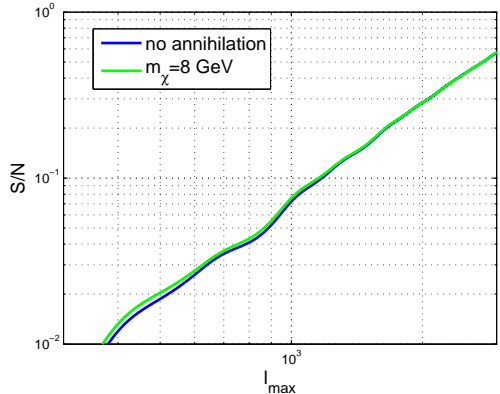}
\caption{Signal-to-noise ratio for a CVL experiment. Blue -- no DM annihilation, green -- $m_\chi=8$~GeV with thermal freezeout annihilation cross section. The plot includes the bispectrum from perturbed visibility only, Eq.~(\ref{eq:full_bispectrum}).}
\label{fig:S2N_Planck}
\end{figure}

The S/N arising from Eq.~(\ref{eq:full_bispectrum}) is plotted in Fig.~\ref{fig:S2N_Planck}, for the case without annihilation and for thermal freezeout annihilation cross section with $m_\chi=8$~GeV. Where S/N is not negligibly small, the annihilation scenario is essentially indistinguishable from the Standard Model. In computing $S/N$, we follow~\cite{Senatore:2008wk} and only include angular scales $l>l_{min}=100$, that are well within the horizon during recombination. This procedure is meant to eliminate the sensitivity of the result to second order metric perturbations, that have not been included in our estimate of the bispectrum. We checked that using smaller values of $l_{min}$ has negligible effect on the results. We consider cosmic variance limited (CVL) experiment with zero detector noise and $f_{\rm sky}=1$\footnote{To compare with expected performance of Planck, there the beam size will cut the growth of $S/N$ above $l\sim1500$.}. 

Lastly, so far we analyzed the bispectrum contribution coming from perturbed visibility, $S_{\delta g}$, but ignored the contribution of the second order source $S^{(2)}$. This is not because the contribution due to $S^{(2)}$ is small; it is in fact comparable to the $S_{\delta g}$ term that we discussed~\cite{Senatore:2008wk}. To our knowledge, a correct analytical estimate for the bispectrum contribution from $S^{(2)}$ is yet to be published. This analysis is motivated regardless of the specific implications for DM and we take it up in~\cite{BDZ}, where we also explain where current estimates~\cite{Senatore:2008wk,Khatri:2009ja} are lacking.  
However, even without the detailed answer for the contribution due to $S^{(2)}$, there is a simple physical argument which makes clear that DM annihilation can only slightly modify this contribution from the Standard Model result. To understand this argument, note that $S^{(2)}$ encodes the cumulative effect of the electron perturbation on the photon multipoles up until last scattering. 

Now, a rather accurate description of CMB anisotropies can be obtained within the tight coupling approximation~\cite{Hu:1995en}, where the effect of Thomson scattering on the  CMB multipoles is packaged into effective diffusion (or Silk) damping,
\ba\label{eq:silk} \Theta_{0,1}(k,\eta)\sim\hat\Theta_{0,1}(k,\eta)\,e^{-\frac{k^2}{k_D^2}}.\ea
Here, $\hat\Theta_{0,1}(k,\eta)$ are the photon monopole and dipole, obtained from the Boltzmann equation deleting Thomson scattering and neglecting all other multipoles, and
\ba \frac{1}{k_D^2(\eta)}=\int_0^\eta d\eta'\frac{c_s^2}{2\dot\tau}\left(\frac{16}{15}+\frac{R^2}{1+R}\right)\ea
is the diffusion scale with $R=(4\rho_b/3\rho_\gamma)$ and $c_s^{-2}=3(1+R)$.

Ref.~\cite{Senatore:2008wk} used Eq.~(\ref{eq:silk}) to derive a rough estimate of the bispectrum contribution due to $S^{(2)}$, limiting the analysis to electron perturbations $\delta_e(k)$ on very large scales, $k\ll k_D$. By construction, their derivation can not strictly apply to small scale $\delta_e$ and thus can not be used to assess quantitatively the effect of DM annihilation. Nevertheless, Ref.~\cite{Senatore:2008wk} pointed out that small scale $\delta_e$ has negligible contribution to the bispectrum, and this observation remains qualitatively correct. The reason to this is simple: electron perturbations on scales smaller than the diffusion mean free path, can not affect diffusion damping. Thus Eq.~(\ref{eq:silk}), by packing Thomson scattering into an effective diffusion coefficient, already implies that short wave electron perturbations cannot lead to big effects. 

We comment that additional contributions associated with $S^{(2)}$ exist, that are not captured by diffusion damping; these contributions are identified in~\cite{BDZ} in terms of perturbations to the photon-baryon sound speed and baryon drag. However, these terms are less significant than the diffusion effect and do not change the results appreciably.

\section{Conclusions}
\label{sec:conc}

We compute linear perturbations to the free electron density $\delta_e$, including the effect of dark matter  (DM) annihilation. We find a growing, non-oscillating, ionization mode that tracks the DM perturbations. The main result of this paper is that on small scales, this growing mode can boost $\delta_e$ by more than an order of magnitude compared to the Standard Model prediction, with peak amplification right after last scattering. The kinetic matter temperature is also affected with potentially $\mathcal{O}(1)$ corrections from the Standard Model prediction during the cosmic dark ages, relevant for 21~cm observations.

CMB power spectra are insensitive to these linear electron density fluctuations. The leading observable where $\delta_e$ may play a role is CMB non-gaussianity, in particular the three-point function or bispectrum. There, a first order electron perturbation feeds into second order, non-gaussian temperature multipoles. Several analytical and numerical studies have shown that the bispectrum from recombination is relevant for Planck and should be accounted for when searching for primordial non-gaussianity. Refs.~\cite{Senatore:2008wk,Khatri:2008kb,Khatri:2009ja} found the bispectrum induced by $\delta_e$ may be marginally observable by Planck. An order of magnitude amplification by DM annihilation then looks naively quite promising; we thus computed the bispectrum induced by $\delta_e$. In doing so, we have found the current literature lacking, specifically when it comes to perturbations on small scales. Our treatment of this problem will be reported separately in~\cite{BDZ}. 

We find that the non-gaussianity signal is small, very difficult to disentangle from the Standard Model by any current or upcoming experiment. This is because even though electron perturbations can be markedly boosted, the main boost occurs slightly after last scattering and on scales below the Silk damping scale. 

While the prospects for observation in the CMB look slim, our analysis does show that significant $\mathcal{O}(1)$ changes to the ionization history of the Universe may be caused by DM interactions during the early cosmic dark ages. In particular, an $\mathcal{O}(1)$ enhancement of electron density and matter temperature perturbations, with power rising on small scales similarly to DM perturbations, would follow from DM annihilation.  Similar conclusions were found for later epochs relating to DM halos; our analysis extends these findings to the early linear regime. A natural observational tool to try and detect these effects in the future is 21~cm radiation. 

\acknowledgments{We thank Tracy Slatyer for early collaboration, and Yacine Ali-Ha\"{i}moud for  comments on the manuscript. CD is supported by the National Science Foundation grant number AST-0807444, NSF Grant number PHY-0855425, and the Raymond and Beverly Sackler Funds. KB is supported by the DOE grant DE-FG02-90ER40542. MZ is supported in part by the National Science Foundation grants PHY-0855425, AST-0907969, PHY-1213563 and by the David \& Lucile Packard.
}
\begin{appendix}

\section{Non-local energy deposition}\label{app:deposition}

Not all of the energy injected by DM annihilation is absorbed by the plasma, and the deposition of the part that is absorbed takes non negligible time. The eventual energy deposition occurs when the annihilation shower becomes an electromagnetic cascade, with electrons and photons cooling down to the $\sim$~keV range where ionization and heating takes over. Here we describe this effect in the homogeneous limit and then proceed to estimate the implications when cosmological perturbations are included. We also make contact with computations of~\cite{Slatyer:2012yq} to illustrate the effect in some concrete model examples.

Ref.~\cite{Slatyer:2012yq} computed the object $T(z,z')$, defined separately for different initial energy $\epsilon_{inj}$ and done for  electrons and photons:
\be T(z,z')\,dz=\frac{dz}{\epsilon_{inj}}\frac{\partial\epsilon}{\partial z}.\ee
This relates to our conventions via
\be f_{dep}(\eta,\eta')=T(z,z')\,H(z).
\ee
In the homogeneous limit, the effect of non-local energy absorption is then encoded by the function $f(\eta)$ of Eq.~(\ref{eq:f}),
\be\label{eq:ffdep} f(\eta)=\int_0^\eta d\eta'\left(a/a'\right)^2\,f_{dep}\left(\eta,\eta'\right)=\int dz\,\frac{H(z)}{H(z')}\,\left(\frac{1+z'}{1+z}\right)^2\,T(z,z').\ee
The case of DM decay, or of time-dependent $\langle\sigma v\rangle$, is a simple generalization of Eqs.~(\ref{eq:f}) and~(\ref{eq:ffdep}).

For the purpose of computing cosmological perturbations, both the time and the spatial smearing of the energy deposition are relevant. Linearizing Eq.~(\ref{eq:smear1}) and moving to Fourier space for clarity,
\ba\label{eq:smear2} \delta\dot u_{dep}(\vec k,\eta)
&=&\dot{u}_{inj}(\eta)\int_0^\eta d\eta'(a/a')^2\left(\delta f_{dep}\left(\vec k,\eta,\eta'\right)+2\,\delta_\chi(\vec k,\eta')\mathcal{F}\left(k,\eta,\eta'\right)\right).\;\;\;\ea
For the homogeneous part of $\mathcal{F}$, we used $\mathcal{F}\left(\vec x,\vec x+\vec r,\eta,\eta'\right)=\mathcal{F}\left(0,|\vec r|,\eta,\eta'\right)$, with Fourier transform $\mathcal{F}\left(k,\eta,\eta'\right)$. For the perturbation in $\mathcal{F}$, it is natural to generalize the quantity $f_{dep}$,
\ba\int d^3x'\delta\mathcal{F}\left(\vec x,\vec x',\eta,\eta'\right)=\delta f_{dep}\left(\vec x,\eta,\eta'\right),\ea
with Fourier transform $\delta f_{dep}(\vec k,\eta,\eta')$.

The term $\delta f_{dep}$ comes from various non-DM perturbations. For instance, electron density perturbations affect the cooling time of energetic photons in the electromagnetic shower following DM annihilation. At late times, $z\lesssim$~few hundreds, baryonic perturbations are as large as DM density perturbations. Then, we expect the $\delta f_{dep}$ term to be as relevant as the DM $\delta_\chi$ term\footnote{This can be relevant e.g. for $21$ cm analyses.}. 
In this paper, however, we restrict our interest to the recombination epoch where, for modes inside the horizon, the DM perturbations $\delta_\chi$ are much larger than all other baryonic (and metric) perturbations. To obtain basic understanding of the physics, it is safe to neglect the $\delta f_{dep}$ term in Eq.~(\ref{eq:smear2}). In addition, again around recombination and for modes inside the horizon $a'\delta_\chi(\vec k,\eta)\approx a\delta_\chi(\vec k,\eta')$. Using these observations we can write,
\ba\label{eq:smear3} \delta\dot u_{dep}(\vec k,\eta)
&\approx&2\,\dot{u}_{inj}(\eta)\,\delta_\chi(\vec k,\eta)\int_0^\eta d\eta'(a/a')\mathcal{F}\left(k,\eta,\eta'\right).
\ea

\begin{figure}[!t]\begin{center}
\includegraphics[width=0.4\textwidth]{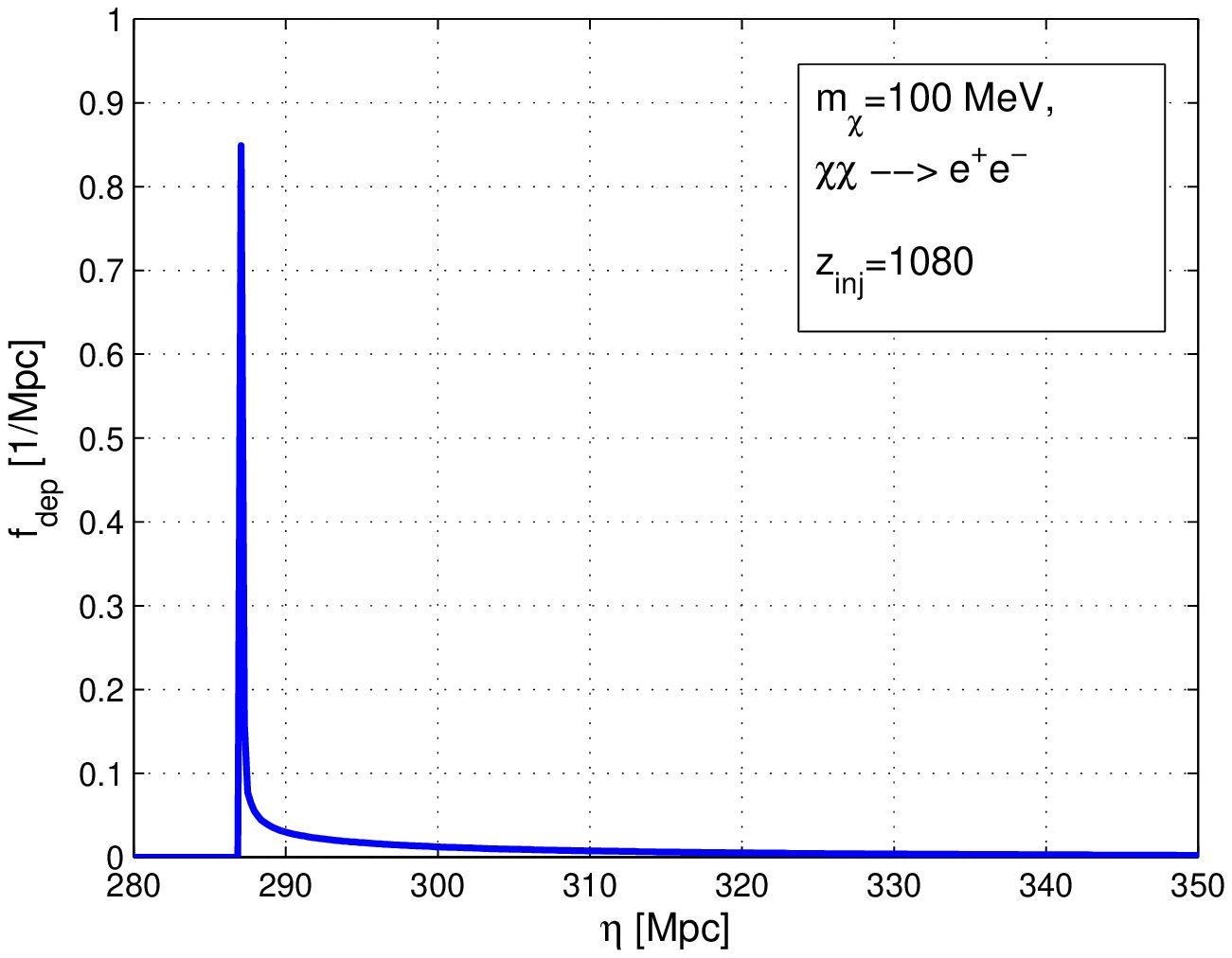}\quad
\includegraphics[width=0.4\textwidth]{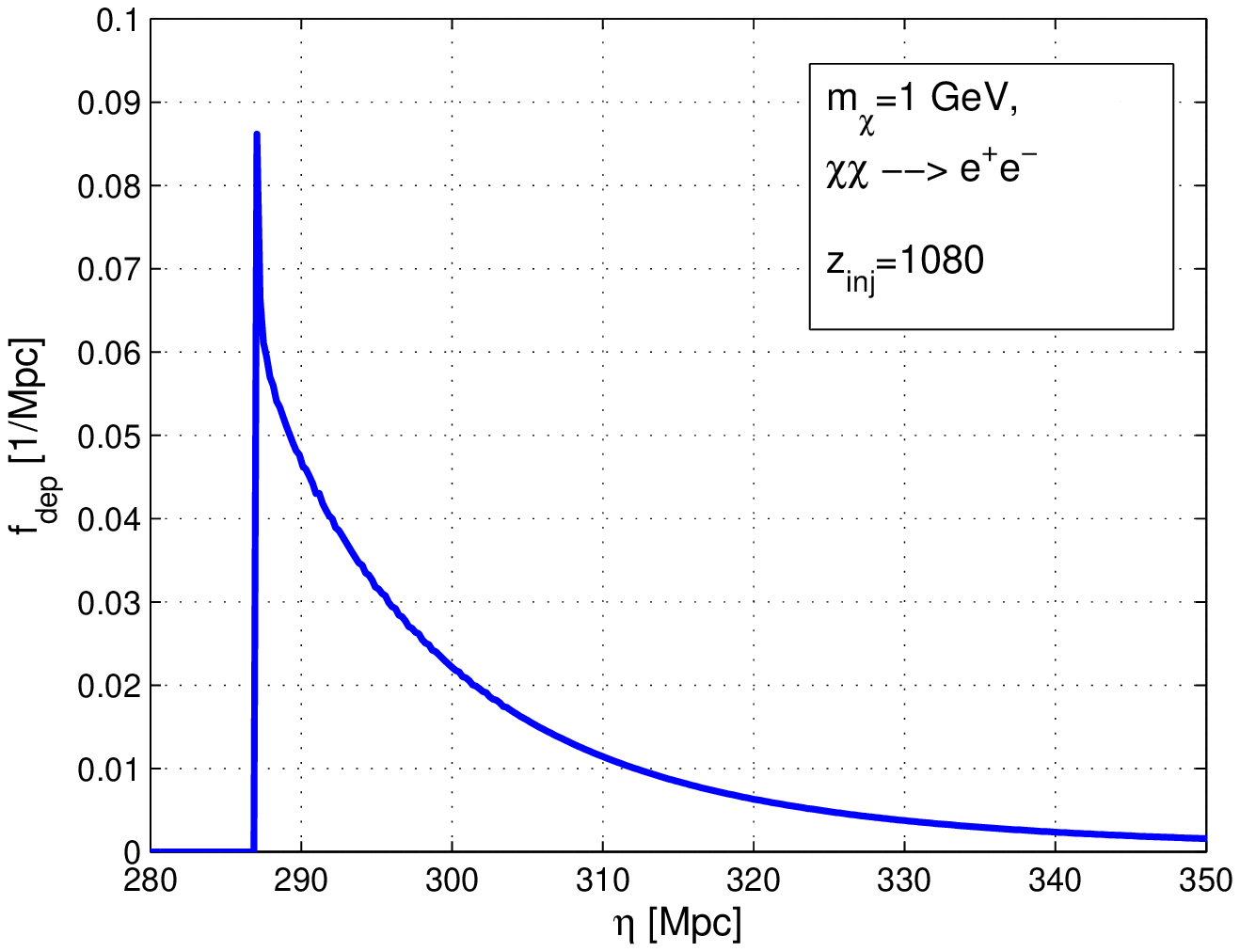}  \end{center}
\caption{Left: energy deposition for $\chi\chi\to e^+e^-$, $m_\chi=100$~MeV, at $z=1080$. Right: same, but for $m_\chi=1$~GeV.}
\label{fig:dep}
\end{figure}

To proceed further, we need information about the model dependent distribution $\mathcal{F}$. Let us consider simple examples. 
\begin{itemize}
\item Instanteneous deposition: consider DM with mass $m_\chi\sim100~$MeV annihilating to $e^+e^-$. Close to the time of recombination at $z\sim10^3$, the electrons cool quickly by inverse Compton (IC) scattering on CMB photons, with a comoving cooling scale
\be k_c\approx2.5\cdot10^3\left(z/10^3\right)^3\left(\epsilon/{\rm GeV}\right)~{\rm Mpc^{-1}}.\ee
For $\sim100$~MeV electrons, this gives $k_c\sim260~$Mpc$^{-1}$ corresponding today to angular resolution $l\sim10^6$, beyond our current ambition. Thus these electrons quickly and locally deliver their energy to IC photons with typical energy $\epsilon_\gamma\sim\gamma_e^2\epsilon_{CMB}\sim$~keV. The ionizing photons quickly deposit their energy in the plasma. For this model it is a reasonable approximation to assume instantaneous deposition,
\be\mathcal{F}\left(\vec x,\vec x',\eta,\eta'\right)\approx\delta^{(3)}(\vec x-\vec x')\delta(\eta-\eta'),\ee
leading to $\delta\dot u_{dep}(\vec x,\eta)\approx2\dot{u}_{inj}(\eta)\delta_\chi(\vec x,\eta)$. 

In the left panel of Fig.~\ref{fig:dep} we explore $f_{dep}(\eta,\eta')$ for the 100~MeV, $\chi\chi\to e^+e^-$ example\footnote{We thank Tracy Slatyer for providing us with high resolution grids of her results of energy deposition.}.
\item Deposition smearing: consider now DM annihilating to $e^+e^-$, but with larger DM mass $m_\chi\sim$~GeV. The initial electrons still cool quickly by IC scattering, however, some fraction of the energy will now go to IC photons with energy $\epsilon_\gamma\sim~$MeV. These MeV photons are non-ionizing; they must cascade down by Thomson scattering to the keV range before they can be absorbed by the plasma. Thus, some fraction $f_{inst}$ of the initial annihilation energy will be deposited locally, but the remaining $1-f_{inst}$ will be smeared over significant distance and time. Consider an ansatz for the deposition smearing,
\ba \mathcal{F}\left(\vec x,\vec x',\eta,\eta'\right)&\approx&f_{inst}(\eta')\delta^{(3)}(\vec x-\vec x')\delta(\eta-\eta')\\
&+&\left(1-f_{inst}(\eta')\right)\left(\frac{1}{\sqrt{2\pi\sigma_\gamma^2(\eta,\eta')}}\right)^3e^{-\frac{\Delta x^2}{2\sigma_\gamma^2(\eta,\eta')}}f_{dep,\gamma}\left(\eta,\eta'\right).\no\ea
Here, $f_{dep,\gamma}$ describes the energy loss rate of the secondary IC photons in the plasma. 
As the photons of interest have MeV energy -- comparable, but not much exceeding the self energy of electrons in the plasma -- the spatial smearing here should be quite similar to CMB diffusion damping. In particular, around the time of recombination we can estimate
\be\sigma^2_\gamma(\eta,\eta')\sim\int_{\eta'}^{\eta}\frac{d\eta''}{3\dot\tau(\eta'')}\sim\frac{4}{k_D^2}.\ee
For the non-local part of the energy deposition, then, we have diffusion damping. Plugging into Eq.~(\ref{eq:smear3}),
\ba\label{eq:smear4}\delta\dot u_{dep}(\vec k,\eta)
&\approx&2\dot{u}_{inj}(\eta)\delta_\chi(\vec k,\eta)\left(f_{inst}(\eta)
+\int_0^\eta d\eta'(a/a')\left(1-f_{inst}(\eta')\right)f_{dep,\gamma}\left(\eta,\eta'\right)e^{-\frac{2k^2}{k_D^2}}\right).\;\;\;\;\;\;\;
\ea

In the right panel of Fig.~\ref{fig:dep} we plot the function $f_{dep}(\eta,\eta')$ for the 1~GeV, $\chi\chi\to e^+e^-$ example. A certain fraction of the injection energy can be attributed to a narrow peak immediately attached to the annihilation time. However, much of the absorbed energy exhibits significant deposition time, much larger than for the previous example and relevant in comparison with the time scale of recombination. This extended deposition can be shown to arise from the shallow behavior of $f_{dep,\gamma}$. 
\end{itemize}
%

To summarize, annihilation energy deposition on small scales is damped by photon diffusion. Considering Eq.~(\ref{eq:smear4}), neglecting the time dependence of all factors compared with that of $f_{dep,\gamma}\left(\eta,\eta'\right)$ in the $\eta'$ integral, we can estimate 
\ba\label{eq:smear5} \delta\dot u_{dep}(\vec k,\eta)\approx2\,\dot{u}_{inj}(\eta)\,\delta_\chi(\vec k,\eta)\left(f_{inst}(\eta)+\left(1-f_{inst}(\eta)\right)\bar f_\gamma(\eta) \,e^{-2k^2/k_D^2}\right),\ea
where $\bar f_\gamma(\eta)=\int d\eta'(a/a')f_{dep,\gamma}\left(\eta,\eta'\right)$. 
In analyzing DM annihilation as a source for cosmological ionization and matter temperature perturbations, we should thus keep in mind that a model dependent, but potentially non-negligible fraction of the annihilation power in DM density perturbations on small scales, $k>k_D\sim0.15$~Mpc$^{-1}$, is washed out and does not source ionization or temperature perturbations on these scales.

Finally it is clear that, analyzing distributions such as in Fig.~\ref{fig:dep}, the quantities $f_{inst}$ and $f_{dep,\gamma}$ (or more generally, the smeared component) can be readily extracted. Once this is done, Eq.~(\ref{eq:smear4}) or~(\ref{eq:smear5})  can be used to calculate the energy absorption damping effect for cosmological perturbations. In this paper, due to the model dependence of the processes involved, we will not go into these details.

\end{appendix}
\bibliography{ref}

\begin{thebibliography}{53}
\expandafter\ifx\csname natexlab\endcsname\relax\def\natexlab#1{#1}\fi
\expandafter\ifx\csname bibnamefont\endcsname\relax
  \def\bibnamefont#1{#1}\fi
\expandafter\ifx\csname bibfnamefont\endcsname\relax
  \def\bibfnamefont#1{#1}\fi
\expandafter\ifx\csname citenamefont\endcsname\relax
  \def\citenamefont#1{#1}\fi
\expandafter\ifx\csname url\endcsname\relax
  \def\url#1{\texttt{#1}}\fi
\expandafter\ifx\csname urlprefix\endcsname\relax\def\urlprefix{URL }\fi
\providecommand{\bibinfo}[2]{#2}
\providecommand{\eprint}[2][]{\url{#2}}

\bibitem[{\citenamefont{Chen and Kamionkowski}(2004)}]{Chen:2003gz}
\bibinfo{author}{\bibfnamefont{X.-L.} \bibnamefont{Chen}} \bibnamefont{and}
  \bibinfo{author}{\bibfnamefont{M.}~\bibnamefont{Kamionkowski}},
  \bibinfo{journal}{Phys.Rev.} \textbf{\bibinfo{volume}{D70}},
  \bibinfo{pages}{043502} (\bibinfo{year}{2004}), \eprint{astro-ph/0310473}.

\bibitem[{\citenamefont{Zhang et~al.}(2006)\citenamefont{Zhang, Chen, Lei, and
  Si}}]{Zhang:2006fr}
\bibinfo{author}{\bibfnamefont{L.}~\bibnamefont{Zhang}},
  \bibinfo{author}{\bibfnamefont{X.-L.} \bibnamefont{Chen}},
  \bibinfo{author}{\bibfnamefont{Y.-A.} \bibnamefont{Lei}}, \bibnamefont{and}
  \bibinfo{author}{\bibfnamefont{Z.-G.} \bibnamefont{Si}},
  \bibinfo{journal}{Phys.Rev.} \textbf{\bibinfo{volume}{D74}},
  \bibinfo{pages}{103519} (\bibinfo{year}{2006}), \eprint{astro-ph/0603425}.

\bibitem[{\citenamefont{Belikov and Hooper}(2009)}]{Belikov:2009qx}
\bibinfo{author}{\bibfnamefont{A.~V.} \bibnamefont{Belikov}} \bibnamefont{and}
  \bibinfo{author}{\bibfnamefont{D.}~\bibnamefont{Hooper}},
  \bibinfo{journal}{Phys.Rev.} \textbf{\bibinfo{volume}{D80}},
  \bibinfo{pages}{035007} (\bibinfo{year}{2009}), \eprint{0904.1210}.

\bibitem[{\citenamefont{Galli et~al.}(2009)\citenamefont{Galli, Iocco, Bertone,
  and Melchiorri}}]{Galli:2009zc}
\bibinfo{author}{\bibfnamefont{S.}~\bibnamefont{Galli}},
  \bibinfo{author}{\bibfnamefont{F.}~\bibnamefont{Iocco}},
  \bibinfo{author}{\bibfnamefont{G.}~\bibnamefont{Bertone}}, \bibnamefont{and}
  \bibinfo{author}{\bibfnamefont{A.}~\bibnamefont{Melchiorri}},
  \bibinfo{journal}{Phys.Rev.} \textbf{\bibinfo{volume}{D80}},
  \bibinfo{pages}{023505} (\bibinfo{year}{2009}), \eprint{0905.0003}.

\bibitem[{\citenamefont{Slatyer et~al.}(2009)\citenamefont{Slatyer,
  Padmanabhan, and Finkbeiner}}]{Slatyer:2009yq}
\bibinfo{author}{\bibfnamefont{T.~R.} \bibnamefont{Slatyer}},
  \bibinfo{author}{\bibfnamefont{N.}~\bibnamefont{Padmanabhan}},
  \bibnamefont{and} \bibinfo{author}{\bibfnamefont{D.~P.}
  \bibnamefont{Finkbeiner}}, \bibinfo{journal}{Phys.Rev.}
  \textbf{\bibinfo{volume}{D80}}, \bibinfo{pages}{043526}
  (\bibinfo{year}{2009}), \eprint{0906.1197}.

\bibitem[{\citenamefont{Cirelli et~al.}(2009)\citenamefont{Cirelli, Iocco, and
  Panci}}]{Cirelli:2009bb}
\bibinfo{author}{\bibfnamefont{M.}~\bibnamefont{Cirelli}},
  \bibinfo{author}{\bibfnamefont{F.}~\bibnamefont{Iocco}}, \bibnamefont{and}
  \bibinfo{author}{\bibfnamefont{P.}~\bibnamefont{Panci}},
  \bibinfo{journal}{JCAP} \textbf{\bibinfo{volume}{0910}}, \bibinfo{pages}{009}
  (\bibinfo{year}{2009}), \eprint{0907.0719}.

\bibitem[{\citenamefont{Natarajan and Schwarz}(2010)}]{Natarajan:2010dc}
\bibinfo{author}{\bibfnamefont{A.}~\bibnamefont{Natarajan}} \bibnamefont{and}
  \bibinfo{author}{\bibfnamefont{D.~J.} \bibnamefont{Schwarz}},
  \bibinfo{journal}{Phys.Rev.} \textbf{\bibinfo{volume}{D81}},
  \bibinfo{pages}{123510} (\bibinfo{year}{2010}), \eprint{1002.4405}.

\bibitem[{\citenamefont{Galli et~al.}(2011)\citenamefont{Galli, Iocco, Bertone,
  and Melchiorri}}]{Galli:2011rz}
\bibinfo{author}{\bibfnamefont{S.}~\bibnamefont{Galli}},
  \bibinfo{author}{\bibfnamefont{F.}~\bibnamefont{Iocco}},
  \bibinfo{author}{\bibfnamefont{G.}~\bibnamefont{Bertone}}, \bibnamefont{and}
  \bibinfo{author}{\bibfnamefont{A.}~\bibnamefont{Melchiorri}},
  \bibinfo{journal}{Phys.Rev.} \textbf{\bibinfo{volume}{D84}},
  \bibinfo{pages}{027302} (\bibinfo{year}{2011}), \eprint{1106.1528}.

\bibitem[{\citenamefont{Finkbeiner et~al.}(2012)\citenamefont{Finkbeiner,
  Galli, Lin, and Slatyer}}]{Finkbeiner:2011dx}
\bibinfo{author}{\bibfnamefont{D.~P.} \bibnamefont{Finkbeiner}},
  \bibinfo{author}{\bibfnamefont{S.}~\bibnamefont{Galli}},
  \bibinfo{author}{\bibfnamefont{T.}~\bibnamefont{Lin}}, \bibnamefont{and}
  \bibinfo{author}{\bibfnamefont{T.~R.} \bibnamefont{Slatyer}},
  \bibinfo{journal}{Phys.Rev.} \textbf{\bibinfo{volume}{D85}},
  \bibinfo{pages}{043522} (\bibinfo{year}{2012}), \eprint{1109.6322}.

\bibitem[{\citenamefont{Hutsi et~al.}(2011)\citenamefont{Hutsi, Chluba, Hektor,
  and Raidal}}]{Hutsi:2011vx}
\bibinfo{author}{\bibfnamefont{G.}~\bibnamefont{Hutsi}},
  \bibinfo{author}{\bibfnamefont{J.}~\bibnamefont{Chluba}},
  \bibinfo{author}{\bibfnamefont{A.}~\bibnamefont{Hektor}}, \bibnamefont{and}
  \bibinfo{author}{\bibfnamefont{M.}~\bibnamefont{Raidal}},
  \bibinfo{journal}{Astron.Astrophys.} \textbf{\bibinfo{volume}{535}},
  \bibinfo{pages}{A26} (\bibinfo{year}{2011}), \eprint{1103.2766}.

\bibitem[{\citenamefont{Natarajan}(2012)}]{Natarajan:2012ry}
\bibinfo{author}{\bibfnamefont{A.}~\bibnamefont{Natarajan}},
  \bibinfo{journal}{Phys.Rev.} \textbf{\bibinfo{volume}{D85}},
  \bibinfo{pages}{083517} (\bibinfo{year}{2012}), \eprint{1201.3939}.

\bibitem[{\citenamefont{Giesen et~al.}(2012)\citenamefont{Giesen, Lesgourgues,
  Audren, and Ali-Haimoud}}]{Giesen:2012rp}
\bibinfo{author}{\bibfnamefont{G.}~\bibnamefont{Giesen}},
  \bibinfo{author}{\bibfnamefont{J.}~\bibnamefont{Lesgourgues}},
  \bibinfo{author}{\bibfnamefont{B.}~\bibnamefont{Audren}}, \bibnamefont{and}
  \bibinfo{author}{\bibfnamefont{Y.}~\bibnamefont{Ali-Haimoud}},
  \bibinfo{journal}{JCAP} \textbf{\bibinfo{volume}{1212}}, \bibinfo{pages}{008}
  (\bibinfo{year}{2012}), \eprint{1209.0247}.

\bibitem[{\citenamefont{Evoli et~al.}(2012{\natexlab{a}})\citenamefont{Evoli,
  Pandolfi, and Ferrara}}]{Evoli:2012qh}
\bibinfo{author}{\bibfnamefont{C.}~\bibnamefont{Evoli}},
  \bibinfo{author}{\bibfnamefont{S.}~\bibnamefont{Pandolfi}}, \bibnamefont{and}
  \bibinfo{author}{\bibfnamefont{A.}~\bibnamefont{Ferrara}}
  (\bibinfo{year}{2012}{\natexlab{a}}), \eprint{1210.6845}.

\bibitem[{\citenamefont{Hinshaw et~al.}(2012)\citenamefont{Hinshaw, Larson,
  Komatsu, Spergel, Bennett et~al.}}]{Hinshaw:2012fq}
\bibinfo{author}{\bibfnamefont{G.}~\bibnamefont{Hinshaw}},
  \bibinfo{author}{\bibfnamefont{D.}~\bibnamefont{Larson}},
  \bibinfo{author}{\bibfnamefont{E.}~\bibnamefont{Komatsu}},
  \bibinfo{author}{\bibfnamefont{D.}~\bibnamefont{Spergel}},
  \bibinfo{author}{\bibfnamefont{C.}~\bibnamefont{Bennett}},
  \bibnamefont{et~al.} (\bibinfo{year}{2012}), \eprint{1212.5226}.

\bibitem[{\citenamefont{Dunkley et~al.}(2013)\citenamefont{Dunkley, Calabrese,
  Sievers, Addison, Battaglia et~al.}}]{Dunkley:2013vu}
\bibinfo{author}{\bibfnamefont{J.}~\bibnamefont{Dunkley}},
  \bibinfo{author}{\bibfnamefont{E.}~\bibnamefont{Calabrese}},
  \bibinfo{author}{\bibfnamefont{J.}~\bibnamefont{Sievers}},
  \bibinfo{author}{\bibfnamefont{G.}~\bibnamefont{Addison}},
  \bibinfo{author}{\bibfnamefont{N.}~\bibnamefont{Battaglia}},
  \bibnamefont{et~al.} (\bibinfo{year}{2013}), \eprint{1301.0776}.

\bibitem[{\citenamefont{Hou et~al.}(2012)\citenamefont{Hou, Reichardt, Story,
  Follin, Keisler et~al.}}]{Hou:2012xq}
\bibinfo{author}{\bibfnamefont{Z.}~\bibnamefont{Hou}},
  \bibinfo{author}{\bibfnamefont{C.}~\bibnamefont{Reichardt}},
  \bibinfo{author}{\bibfnamefont{K.}~\bibnamefont{Story}},
  \bibinfo{author}{\bibfnamefont{B.}~\bibnamefont{Follin}},
  \bibinfo{author}{\bibfnamefont{R.}~\bibnamefont{Keisler}},
  \bibnamefont{et~al.} (\bibinfo{year}{2012}), \eprint{1212.6267}.

\bibitem[{\citenamefont{Slatyer}(2012)}]{Slatyer:2012yq}
\bibinfo{author}{\bibfnamefont{T.~R.} \bibnamefont{Slatyer}}
  (\bibinfo{year}{2012}), \eprint{1211.0283}.

\bibitem[{\citenamefont{Senatore
  et~al.}(2009{\natexlab{a}})\citenamefont{Senatore, Tassev, and
  Zaldarriaga}}]{Senatore:2008vi}
\bibinfo{author}{\bibfnamefont{L.}~\bibnamefont{Senatore}},
  \bibinfo{author}{\bibfnamefont{S.}~\bibnamefont{Tassev}}, \bibnamefont{and}
  \bibinfo{author}{\bibfnamefont{M.}~\bibnamefont{Zaldarriaga}},
  \bibinfo{journal}{JCAP} \textbf{\bibinfo{volume}{0908}}, \bibinfo{pages}{031}
  (\bibinfo{year}{2009}{\natexlab{a}}), \eprint{0812.3652}.

\bibitem[{\citenamefont{Novosyadlyj}(2006)}]{Novosyadlyj:2006fw}
\bibinfo{author}{\bibfnamefont{B.}~\bibnamefont{Novosyadlyj}},
  \bibinfo{journal}{Mon.Not.Roy.Astron.Soc.} \textbf{\bibinfo{volume}{370}},
  \bibinfo{pages}{1771} (\bibinfo{year}{2006}), \eprint{astro-ph/0603674}.

\bibitem[{\citenamefont{Venhlovska and Novosyadlyj}(2008)}]{Venhlovska:2008uc}
\bibinfo{author}{\bibfnamefont{B.}~\bibnamefont{Venhlovska}} \bibnamefont{and}
  \bibinfo{author}{\bibfnamefont{B.}~\bibnamefont{Novosyadlyj}},
  \bibinfo{journal}{J.Phys.Stud.} \textbf{\bibinfo{volume}{12}},
  \bibinfo{pages}{3901} (\bibinfo{year}{2008}), \eprint{0812.2452}.

\bibitem[{\citenamefont{Furlanetto
  et~al.}(2006{\natexlab{a}})\citenamefont{Furlanetto, Oh, and
  Pierpaoli}}]{Furlanetto:2006wp}
\bibinfo{author}{\bibfnamefont{S.~R.} \bibnamefont{Furlanetto}},
  \bibinfo{author}{\bibfnamefont{S.~P.} \bibnamefont{Oh}}, \bibnamefont{and}
  \bibinfo{author}{\bibfnamefont{E.}~\bibnamefont{Pierpaoli}},
  \bibinfo{journal}{Phys.Rev.} \textbf{\bibinfo{volume}{D74}},
  \bibinfo{pages}{103502} (\bibinfo{year}{2006}{\natexlab{a}}),
  \eprint{astro-ph/0608385}.

\bibitem[{\citenamefont{Senatore
  et~al.}(2009{\natexlab{b}})\citenamefont{Senatore, Tassev, and
  Zaldarriaga}}]{Senatore:2008wk}
\bibinfo{author}{\bibfnamefont{L.}~\bibnamefont{Senatore}},
  \bibinfo{author}{\bibfnamefont{S.}~\bibnamefont{Tassev}}, \bibnamefont{and}
  \bibinfo{author}{\bibfnamefont{M.}~\bibnamefont{Zaldarriaga}},
  \bibinfo{journal}{JCAP} \textbf{\bibinfo{volume}{0909}}, \bibinfo{pages}{038}
  (\bibinfo{year}{2009}{\natexlab{b}}), \eprint{0812.3658}.

\bibitem[{\citenamefont{Khatri and Wandelt}(2009)}]{Khatri:2008kb}
\bibinfo{author}{\bibfnamefont{R.}~\bibnamefont{Khatri}} \bibnamefont{and}
  \bibinfo{author}{\bibfnamefont{B.~D.} \bibnamefont{Wandelt}},
  \bibinfo{journal}{Phys.Rev.} \textbf{\bibinfo{volume}{D79}},
  \bibinfo{pages}{023501} (\bibinfo{year}{2009}), \eprint{0810.4370}.

\bibitem[{\citenamefont{Khatri and Wandelt}(2010)}]{Khatri:2009ja}
\bibinfo{author}{\bibfnamefont{R.}~\bibnamefont{Khatri}} \bibnamefont{and}
  \bibinfo{author}{\bibfnamefont{B.~D.} \bibnamefont{Wandelt}},
  \bibinfo{journal}{Phys.Rev.} \textbf{\bibinfo{volume}{D81}},
  \bibinfo{pages}{103518} (\bibinfo{year}{2010}), \eprint{0903.0871}.

\bibitem[{\citenamefont{Pitrou et~al.}(2010)\citenamefont{Pitrou, Uzan, and
  Bernardeau}}]{Pitrou:2010sn}
\bibinfo{author}{\bibfnamefont{C.}~\bibnamefont{Pitrou}},
  \bibinfo{author}{\bibfnamefont{J.-P.} \bibnamefont{Uzan}}, \bibnamefont{and}
  \bibinfo{author}{\bibfnamefont{F.}~\bibnamefont{Bernardeau}},
  \bibinfo{journal}{JCAP} \textbf{\bibinfo{volume}{1007}}, \bibinfo{pages}{003}
  (\bibinfo{year}{2010}), \eprint{1003.0481}.

\bibitem[{\citenamefont{Huang and Vernizzi}(2012)}]{Huang:2012ub}
\bibinfo{author}{\bibfnamefont{Z.}~\bibnamefont{Huang}} \bibnamefont{and}
  \bibinfo{author}{\bibfnamefont{F.}~\bibnamefont{Vernizzi}}
  (\bibinfo{year}{2012}), \eprint{1212.3573}.

\bibitem[{\citenamefont{Su et~al.}(2012)\citenamefont{Su, Lim, and
  Shellard}}]{Su:2012gt}
\bibinfo{author}{\bibfnamefont{S.-C.} \bibnamefont{Su}},
  \bibinfo{author}{\bibfnamefont{E.~A.} \bibnamefont{Lim}}, \bibnamefont{and}
  \bibinfo{author}{\bibfnamefont{E.}~\bibnamefont{Shellard}}
  (\bibinfo{year}{2012}), \eprint{1212.6968}.

\bibitem[{\citenamefont{Pettinari et~al.}(2013)\citenamefont{Pettinari, Fidler,
  Crittenden, Koyama, and Wands}}]{Pettinari:2013he}
\bibinfo{author}{\bibfnamefont{G.~W.} \bibnamefont{Pettinari}},
  \bibinfo{author}{\bibfnamefont{C.}~\bibnamefont{Fidler}},
  \bibinfo{author}{\bibfnamefont{R.}~\bibnamefont{Crittenden}},
  \bibinfo{author}{\bibfnamefont{K.}~\bibnamefont{Koyama}}, \bibnamefont{and}
  \bibinfo{author}{\bibfnamefont{D.}~\bibnamefont{Wands}}
  (\bibinfo{year}{2013}), \eprint{1302.0832}.

\bibitem[{\citenamefont{Blum et~al.}(2013)\citenamefont{Blum, Dvorkin, and
  Zaldarriaga}}]{BDZ}
\bibinfo{author}{\bibfnamefont{K.}~\bibnamefont{Blum}},
  \bibinfo{author}{\bibfnamefont{C.}~\bibnamefont{Dvorkin}}, \bibnamefont{and}
  \bibinfo{author}{\bibfnamefont{M.}~\bibnamefont{Zaldarriaga}},
  \bibinfo{journal}{To appear}  (\bibinfo{year}{2013}).

\bibitem[{\citenamefont{Valdes et~al.}(2007)\citenamefont{Valdes, Ferrara,
  Mapelli, and Ripamonti}}]{Valdes:2007cu}
\bibinfo{author}{\bibfnamefont{M.}~\bibnamefont{Valdes}},
  \bibinfo{author}{\bibfnamefont{A.}~\bibnamefont{Ferrara}},
  \bibinfo{author}{\bibfnamefont{M.}~\bibnamefont{Mapelli}}, \bibnamefont{and}
  \bibinfo{author}{\bibfnamefont{E.}~\bibnamefont{Ripamonti}},
  \bibinfo{journal}{Mon.Not.Roy.Astron.Soc.} \textbf{\bibinfo{volume}{377}},
  \bibinfo{pages}{245} (\bibinfo{year}{2007}), \eprint{astro-ph/0701301}.

\bibitem[{\citenamefont{Cumberbatch et~al.}(2010)\citenamefont{Cumberbatch,
  Lattanzi, Silk, Lattanzi, and Silk}}]{Cumberbatch:2008rh}
\bibinfo{author}{\bibfnamefont{D.~T.} \bibnamefont{Cumberbatch}},
  \bibinfo{author}{\bibfnamefont{M.}~\bibnamefont{Lattanzi}},
  \bibinfo{author}{\bibfnamefont{J.}~\bibnamefont{Silk}},
  \bibinfo{author}{\bibfnamefont{M.}~\bibnamefont{Lattanzi}}, \bibnamefont{and}
  \bibinfo{author}{\bibfnamefont{J.}~\bibnamefont{Silk}},
  \bibinfo{journal}{Phys.Rev.} \textbf{\bibinfo{volume}{D82}},
  \bibinfo{pages}{103508} (\bibinfo{year}{2010}), \eprint{0808.0881}.

\bibitem[{\citenamefont{Finkbeiner et~al.}(2008)\citenamefont{Finkbeiner,
  Padmanabhan, and Weiner}}]{Finkbeiner:2008gw}
\bibinfo{author}{\bibfnamefont{D.~P.} \bibnamefont{Finkbeiner}},
  \bibinfo{author}{\bibfnamefont{N.}~\bibnamefont{Padmanabhan}},
  \bibnamefont{and} \bibinfo{author}{\bibfnamefont{N.}~\bibnamefont{Weiner}},
  \bibinfo{journal}{Phys.Rev.} \textbf{\bibinfo{volume}{D78}},
  \bibinfo{pages}{063530} (\bibinfo{year}{2008}), \eprint{0805.3531}.

\bibitem[{\citenamefont{Natarajan and Schwarz}(2009)}]{Natarajan:2009bm}
\bibinfo{author}{\bibfnamefont{A.}~\bibnamefont{Natarajan}} \bibnamefont{and}
  \bibinfo{author}{\bibfnamefont{D.~J.} \bibnamefont{Schwarz}},
  \bibinfo{journal}{Phys.Rev.} \textbf{\bibinfo{volume}{D80}},
  \bibinfo{pages}{043529} (\bibinfo{year}{2009}), \eprint{0903.4485}.

\bibitem[{\citenamefont{Valdes et~al.}(2012)\citenamefont{Valdes, Evoli,
  Mesinger, Ferrara, and Yoshida}}]{Valdes:2012zv}
\bibinfo{author}{\bibfnamefont{M.}~\bibnamefont{Valdes}},
  \bibinfo{author}{\bibfnamefont{C.}~\bibnamefont{Evoli}},
  \bibinfo{author}{\bibfnamefont{A.}~\bibnamefont{Mesinger}},
  \bibinfo{author}{\bibfnamefont{A.}~\bibnamefont{Ferrara}}, \bibnamefont{and}
  \bibinfo{author}{\bibfnamefont{N.}~\bibnamefont{Yoshida}}
  (\bibinfo{year}{2012}), \eprint{1209.2120}.

\bibitem[{\citenamefont{Komatsu et~al.}(2011)}]{Komatsu:2010fb}
\bibinfo{author}{\bibfnamefont{E.}~\bibnamefont{Komatsu}} \bibnamefont{et~al.}
  (\bibinfo{collaboration}{WMAP Collaboration}),
  \bibinfo{journal}{Astrophys.J.Suppl.} \textbf{\bibinfo{volume}{192}},
  \bibinfo{pages}{18} (\bibinfo{year}{2011}), \eprint{1001.4538}.

\bibitem[{\citenamefont{{Peebles}}(1968)}]{1968ApJPeebles}
\bibinfo{author}{\bibfnamefont{P.~J.~E.} \bibnamefont{{Peebles}}},
  \bibinfo{journal}{\apj} \textbf{\bibinfo{volume}{153}}, \bibinfo{pages}{1}
  (\bibinfo{year}{1968}).

\bibitem[{\citenamefont{Y.~B.~Zeldovich and Sunyaev}(1969)}]{Zeldovich}
\bibinfo{author}{\bibfnamefont{V.~G.~K.} \bibnamefont{Y.~B.~Zeldovich}}
  \bibnamefont{and} \bibinfo{author}{\bibfnamefont{R.~A.}
  \bibnamefont{Sunyaev}}, \bibinfo{journal}{J. Exp. Theor. Phys.}
  \textbf{\bibinfo{volume}{28}}, \bibinfo{pages}{146} (\bibinfo{year}{1969}).

\bibitem[{\citenamefont{Seager et~al.}(1999)\citenamefont{Seager, Sasselov, and
  Scott}}]{Seager:1999bc}
\bibinfo{author}{\bibfnamefont{S.}~\bibnamefont{Seager}},
  \bibinfo{author}{\bibfnamefont{D.~D.} \bibnamefont{Sasselov}},
  \bibnamefont{and} \bibinfo{author}{\bibfnamefont{D.}~\bibnamefont{Scott}},
  \bibinfo{journal}{Astrophys.J.} \textbf{\bibinfo{volume}{523}},
  \bibinfo{pages}{L1} (\bibinfo{year}{1999}), \eprint{astro-ph/9909275}.

\bibitem[{\citenamefont{Seager et~al.}(2000)\citenamefont{Seager, Sasselov, and
  Scott}}]{Seager:1999km}
\bibinfo{author}{\bibfnamefont{S.}~\bibnamefont{Seager}},
  \bibinfo{author}{\bibfnamefont{D.~D.} \bibnamefont{Sasselov}},
  \bibnamefont{and} \bibinfo{author}{\bibfnamefont{D.}~\bibnamefont{Scott}},
  \bibinfo{journal}{Astrophys.J.Suppl.} \textbf{\bibinfo{volume}{128}},
  \bibinfo{pages}{407} (\bibinfo{year}{2000}), \eprint{astro-ph/9912182}.

\bibitem[{\citenamefont{Valdes et~al.}(2010)\citenamefont{Valdes, Evoli, and
  Ferrara}}]{Valdes:2009cq}
\bibinfo{author}{\bibfnamefont{M.}~\bibnamefont{Valdes}},
  \bibinfo{author}{\bibfnamefont{C.}~\bibnamefont{Evoli}}, \bibnamefont{and}
  \bibinfo{author}{\bibfnamefont{A.}~\bibnamefont{Ferrara}},
  \bibinfo{journal}{Mon.Not.Roy.Astron.Soc.} \textbf{\bibinfo{volume}{404}},
  \bibinfo{pages}{1569} (\bibinfo{year}{2010}), \eprint{0911.1125}.

\bibitem[{\citenamefont{Evoli et~al.}(2012{\natexlab{b}})\citenamefont{Evoli,
  Valdes, Ferrara, and Yoshida}}]{Evoli:2012zz}
\bibinfo{author}{\bibfnamefont{C.}~\bibnamefont{Evoli}},
  \bibinfo{author}{\bibfnamefont{M.}~\bibnamefont{Valdes}},
  \bibinfo{author}{\bibfnamefont{A.}~\bibnamefont{Ferrara}}, \bibnamefont{and}
  \bibinfo{author}{\bibfnamefont{N.}~\bibnamefont{Yoshida}},
  \bibinfo{journal}{Mon.Not.Roy.Astron.Soc.} \textbf{\bibinfo{volume}{422}},
  \bibinfo{pages}{420} (\bibinfo{year}{2012}{\natexlab{b}}).

\bibitem[{\citenamefont{Padmanabhan and Finkbeiner}(2005)}]{Padmanabhan:2005es}
\bibinfo{author}{\bibfnamefont{N.}~\bibnamefont{Padmanabhan}} \bibnamefont{and}
  \bibinfo{author}{\bibfnamefont{D.~P.} \bibnamefont{Finkbeiner}},
  \bibinfo{journal}{Phys.Rev.} \textbf{\bibinfo{volume}{D72}},
  \bibinfo{pages}{023508} (\bibinfo{year}{2005}), \eprint{astro-ph/0503486}.

\bibitem[{\citenamefont{Switzer and Hirata}(2008)}]{Switzer:2007sn}
\bibinfo{author}{\bibfnamefont{E.~R.} \bibnamefont{Switzer}} \bibnamefont{and}
  \bibinfo{author}{\bibfnamefont{C.~M.} \bibnamefont{Hirata}},
  \bibinfo{journal}{Phys.Rev.} \textbf{\bibinfo{volume}{D77}},
  \bibinfo{pages}{083006} (\bibinfo{year}{2008}), \eprint{astro-ph/0702143}.

\bibitem[{\citenamefont{Ma and Bertschinger}(1995)}]{Ma:1995ey}
\bibinfo{author}{\bibfnamefont{C.-P.} \bibnamefont{Ma}} \bibnamefont{and}
  \bibinfo{author}{\bibfnamefont{E.}~\bibnamefont{Bertschinger}},
  \bibinfo{journal}{Astrophys.J.} \textbf{\bibinfo{volume}{455}},
  \bibinfo{pages}{7} (\bibinfo{year}{1995}), \eprint{astro-ph/9506072}.

\bibitem[{\citenamefont{Lewis}()}]{camb_notes}
\bibinfo{author}{\bibfnamefont{A.}~\bibnamefont{Lewis}},
  \bibinfo{note}{\url{http://cosmologist.info/notes/CAMB.pdf}}.

\bibitem[{\citenamefont{Lewis et~al.}(2000)\citenamefont{Lewis, Challinor, and
  Lasenby}}]{Lewis:1999bs}
\bibinfo{author}{\bibfnamefont{A.}~\bibnamefont{Lewis}},
  \bibinfo{author}{\bibfnamefont{A.}~\bibnamefont{Challinor}},
  \bibnamefont{and} \bibinfo{author}{\bibfnamefont{A.}~\bibnamefont{Lasenby}},
  \bibinfo{journal}{Astrophys. J.} \textbf{\bibinfo{volume}{538}},
  \bibinfo{pages}{473} (\bibinfo{year}{2000}), \eprint{astro-ph/9911177}.

\bibitem[{\citenamefont{Howlett et~al.}(2012)\citenamefont{Howlett, Lewis,
  Hall, and Challinor}}]{Howlett:2012mh}
\bibinfo{author}{\bibfnamefont{C.}~\bibnamefont{Howlett}},
  \bibinfo{author}{\bibfnamefont{A.}~\bibnamefont{Lewis}},
  \bibinfo{author}{\bibfnamefont{A.}~\bibnamefont{Hall}}, \bibnamefont{and}
  \bibinfo{author}{\bibfnamefont{A.}~\bibnamefont{Challinor}},
  \bibinfo{journal}{JCAP} \textbf{\bibinfo{volume}{1204}}, \bibinfo{pages}{027}
  (\bibinfo{year}{2012}), \eprint{1201.3654}.

\bibitem[{\citenamefont{Furlanetto
  et~al.}(2006{\natexlab{b}})\citenamefont{Furlanetto, Oh, and
  Briggs}}]{Furlanetto:2006jb}
\bibinfo{author}{\bibfnamefont{S.}~\bibnamefont{Furlanetto}},
  \bibinfo{author}{\bibfnamefont{S.~P.} \bibnamefont{Oh}}, \bibnamefont{and}
  \bibinfo{author}{\bibfnamefont{F.}~\bibnamefont{Briggs}},
  \bibinfo{journal}{Phys.Rept.} \textbf{\bibinfo{volume}{433}},
  \bibinfo{pages}{181} (\bibinfo{year}{2006}{\natexlab{b}}),
  \eprint{astro-ph/0608032}.

\bibitem[{\citenamefont{Loeb and Zaldarriaga}(2004)}]{Loeb:2003ya}
\bibinfo{author}{\bibfnamefont{A.}~\bibnamefont{Loeb}} \bibnamefont{and}
  \bibinfo{author}{\bibfnamefont{M.}~\bibnamefont{Zaldarriaga}},
  \bibinfo{journal}{Phys.Rev.Lett.} \textbf{\bibinfo{volume}{92}},
  \bibinfo{pages}{211301} (\bibinfo{year}{2004}), \eprint{astro-ph/0312134}.

\bibitem[{\citenamefont{Lewis and Challinor}(2007)}]{Lewis:2007kz}
\bibinfo{author}{\bibfnamefont{A.}~\bibnamefont{Lewis}} \bibnamefont{and}
  \bibinfo{author}{\bibfnamefont{A.}~\bibnamefont{Challinor}},
  \bibinfo{journal}{Phys.Rev.} \textbf{\bibinfo{volume}{D76}},
  \bibinfo{pages}{083005} (\bibinfo{year}{2007}), \eprint{astro-ph/0702600}.

\bibitem[{\citenamefont{Zaldarriaga and Seljak}(1997)}]{Zaldarriaga:1996xe}
\bibinfo{author}{\bibfnamefont{M.}~\bibnamefont{Zaldarriaga}} \bibnamefont{and}
  \bibinfo{author}{\bibfnamefont{U.}~\bibnamefont{Seljak}},
  \bibinfo{journal}{Phys.Rev.} \textbf{\bibinfo{volume}{D55}},
  \bibinfo{pages}{1830} (\bibinfo{year}{1997}), \eprint{astro-ph/9609170}.

\bibitem[{\citenamefont{Komatsu and Spergel}(2001)}]{Komatsu:2001rj}
\bibinfo{author}{\bibfnamefont{E.}~\bibnamefont{Komatsu}} \bibnamefont{and}
  \bibinfo{author}{\bibfnamefont{D.~N.} \bibnamefont{Spergel}},
  \bibinfo{journal}{Phys.Rev.} \textbf{\bibinfo{volume}{D63}},
  \bibinfo{pages}{063002} (\bibinfo{year}{2001}), \eprint{astro-ph/0005036}.

\bibitem[{\citenamefont{Hu and Sugiyama}(1996)}]{Hu:1995en}
\bibinfo{author}{\bibfnamefont{W.}~\bibnamefont{Hu}} \bibnamefont{and}
  \bibinfo{author}{\bibfnamefont{N.}~\bibnamefont{Sugiyama}},
  \bibinfo{journal}{Astrophys.J.} \textbf{\bibinfo{volume}{471}},
  \bibinfo{pages}{542} (\bibinfo{year}{1996}), \eprint{astro-ph/9510117}.

\end{thebibliography}

\end{document}